\documentclass[iop, numberedappendix]{emulateapj}
\usepackage{amssymb}
\usepackage{amsmath}
\usepackage{float}
\usepackage{relsize}
\usepackage{verbatim}
\usepackage{color}
\usepackage{comment}

\def\be{\begin{equation}}
\def\ee{\end{equation}}
\def\ba{\begin{eqnarray}}
\def\ea{\end{eqnarray}}

\newcommand{\DM}{\ensuremath{\mathrm{DM}}} 
\newcommand{\RM}{\ensuremath{\mathrm{RM}}} 
\newcommand{\SN}{\ensuremath{\mathrm{S/N}}}
\newcommand{\Smin}{\ensuremath{S_{\mathrm{min}}}}
\newcommand{\tsub}{\ensuremath{t_{\mathrm{sub}}}}
\newcommand{\J}{\ensuremath{\mathrm{J}}}
\newcommand{\C}{\ensuremath{\mathrm{C}}}
\newcommand{\DISS}{\ensuremath{\mathrm{DISS}}}
\newcommand{\RISS}{\ensuremath{\mathrm{RISS}}}
\newcommand{\PBF}{\ensuremath{\mathrm{PBF}}}
\newcommand{\CCF}{\ensuremath{\mathrm{CCF}}}
\newcommand{\Like}{\ensuremath{\mathcal{L}}}
\newcommand{\R}{\ensuremath{\mathcal{R}}}
\newcommand{\Nphi}{\ensuremath{N_\phi}}
\newcommand{\Np}{\ensuremath{N_\mathrm{p}}}
\newcommand{\fJ}{\ensuremath{f_\mathrm{J}}}
\newcommand{\kJ}{\ensuremath{k_\mathrm{J}}}
\newcommand{\kJc}{\ensuremath{k_\mathrm{J,c}}}

\newcommand{\mIc}{\ensuremath{m_\mathrm{I,c}}}
\newcommand{\Weff}{\ensuremath{W_\mathrm{eff}}}
\newcommand{\half}{\ensuremath{\frac{1}{2}}}
\newcommand{\cmthree}{\ensuremath{\mathrm{cm^{-3}}}}

\newcommand{\niss}{n_{\rm ISS}}

\newcommand{\Dt}{\Delta t}
\newcommand{\Dto}{\Delta t_0}

\newcommand{\Dtd}{\Delta t_{\mathrm{d}}}
\newcommand{\Dnud}{\Delta \nu_{\mathrm{d}}}
\newcommand{\taud}{\tau_{\mathrm{d}}}

\shorttitle{Noise Budget for Pulsar Timing}
\shortauthors{Lam et al.}

\begin{document}

\title{
The NANOGrav Nine-Year Data Set:
 Noise Budget for Pulsar Arrival Times on Intraday Timescales
}
\author{ 
M.\,T.\,Lam\altaffilmark{1},
J.\,M.\,Cordes\altaffilmark{1},
S.\,Chatterjee\altaffilmark{1}, 
Z.\,Arzoumanian\altaffilmark{2},
K.\,Crowter\altaffilmark{3},
P.\,B.\,Demorest\altaffilmark{4},
T.\,Dolch\altaffilmark{1,5},
J.\,A.\,Ellis\altaffilmark{6,19},
R.\,D.\,Ferdman\altaffilmark{7},
E.\,F.\,Fonseca\altaffilmark{3},
M.\,E.\,Gonzalez\altaffilmark{3,8},
G.\,Jones\altaffilmark{9},
M.\,L.\,Jones\altaffilmark{10},
L.\,Levin\altaffilmark{10,11},
D.\,R.\,Madison\altaffilmark{1,12},
M.\,A.\,McLaughlin\altaffilmark{10},
D.\,J.\,Nice\altaffilmark{13},
T.\,T.\,Pennucci\altaffilmark{14,15},
S.\,M.\,Ransom\altaffilmark{12},
X.\,Siemens\altaffilmark{16},
I.\,H.\,Stairs\altaffilmark{3},
K.\,Stovall\altaffilmark{17},
J.\,K.\,Swiggum\altaffilmark{10,16},
W.\,W.\,Zhu\altaffilmark{3,18}}
\altaffiltext{1}{Department of Astronomy and Cornell Center for Astrophysics and Planetary Science, Cornell University, Ithaca, NY 14853, USA; mlam@astro.cornell.edu}
\altaffiltext{2}{Center for Research and Exploration in Space Science and Technology and X-Ray Astrophysics Laboratory, NASA Goddard Space Flight Center, Code 662, Greenbelt, MD 20771, USA}
\altaffiltext{3}{Department of Physics and Astronomy, University of British Columbia, 6224 Agricultural Road, Vancouver, BC V6T 1Z1, Canada}
\altaffiltext{4}{National Radio Astronomy Observatory, P.O.~Box 0, Socorro, NM, 87801, USA}
\altaffiltext{5}{Department of Physics, Hillsdale College, 33 E. College Street, Hillsdale, MI 49242, USA}
\altaffiltext{6}{Jet Propulsion Laboratory, California Institute of Technology, 4800 Oak Grove Dr. Pasadena CA, 91109, USA}
\altaffiltext{7}{Department of Physics, McGill University, 3600 rue Universite, Montreal, QC H3A 2T8, Canada}
\altaffiltext{8}{Department of Nuclear Medicine, Vancouver Coastal Health Authority, Vancouver, BC V5Z 1M9, Canada}
\altaffiltext{9}{Department of Physics, Columbia University, 550 W. 120th St. New York, NY 10027, USA}
\altaffiltext{10}{Department of Physics, West Virginia University, P.O.~Box 6315, Morgantown, WV 26505, USA}

\begin{abstract}

The use of pulsars as astrophysical clocks for gravitational wave experiments demands the highest possible timing precision. Pulse times of arrival (TOAs) are limited by stochastic processes that occur in the pulsar itself, along the line of sight through the interstellar medium, and in the measurement process. On timescales of seconds to hours, the TOA variance exceeds that from template-fitting errors due to additive noise. We assess contributions to the total variance from two additional effects: amplitude and phase jitter intrinsic to single pulses and changes in the interstellar impulse response from scattering. The three effects have different dependencies on time, frequency, and pulse signal-to-noise ratio. We use data on 37 pulsars from the North American Nanohertz Observatory for Gravitational Waves to assess the individual contributions to the overall intraday noise budget for each pulsar. We detect jitter in 22 pulsars and estimate the average value of RMS jitter in our pulsars to be $\sim 1\%$ of pulse phase. We examine how jitter evolves as a function of frequency and find evidence for evolution. Finally, we compare our measurements with previous noise parameter estimates and discuss methods to improve gravitational wave detection pipelines.

\end{abstract}


\section{Introduction}

Pulsar timing is used for a variety of unique applications in astrophysics and fundamental physics. These include mass determinations of neutron stars (NSs) and their binary companions to contrain compact object formation mechanisms and equations-of-state \citep{Demorest+2010,Antoniadis}; precision tests of general relativity and other theories of gravity \citep{Will2014}; limits on changes in fundamental constants \citep{Lazaridis+2009,sw2013,Zhu+2015}; and, especially recently,  using arrays of pulsars as detectors of low-frequency (nanohertz) gravitational waves (GWs; e.g., \citealt{NG5BWM,NG9GWB}).  Improvements  in the accuracy of measured arrival times continue to yield benefits in these applications.  In this paper, we present a detailed assessment of the time-of-arrival (TOA) noise budget that is applicable to measurements made on relatively short timescales, ranging from single pulse periods to  integration times of 10 -- $10^4$~s.   The work discussed here complements other studies that address noise contributions from variations in  the spin rates of neutron stars \citep[e.g.,][]{2010MNRAS.402.1027H, sc2010},  the frequency dependence of pulse shapes \citep[][]{pdr2014}, and from propagation through the interstellar medium \citep[ISM;][]{1984Natur.307..527A, 1984JApA....5..369B, r90, fc1990, cs2010}. 

\footnotetext[11]{Jodrell Bank Centre for Astrophysics, School of Physics and Astronomy, The University of Manchester, Manchester M13 9PL, UK}
\footnotetext[12]{National Radio Astronomy Observatory, 520 Edgemont Road, Charlottesville, VA 22903, USA}
\footnotetext[13]{Department of Physics, Lafayette College, Easton, PA 18042, USA}
\footnotetext[14]{University of Virginia, Department of Astronomy, P.O.~Box 400325, Charlottesville, VA 22904-4325, USA}
\footnotetext[15]{Department of Physics, Columbia University, 550 W. 120th St. New York, NY 10027, USA}
\footnotetext[16]{Center for Gravitation, Cosmology and Astrophysics, Department of Physics, University of Wisconsin-Milwaukee, P.O. Box 413, Milwaukee, WI 53201, USA}
\footnotetext[17]{Department of Physics and Astronomy, University of New Mexico, Albuquerque, NM, 87131, USA}
\footnotetext[18]{Max-Planck-Institut f\"{u}r Radioastronomie, Auf dem H\"{u}gel 69, D-53121, Bonn, Germany}
\footnotetext[19]{Einstein Fellow}

\setcounter{footnote}{0}

Pulsar timing relies on a foundation of pulsar phenomena that have been demonstrated over the nearly half century since pulsars were discovered (see \citealt{Cordes2013} for a review). Rotational stability, especially for recycled millisecond pulsars (MSPs), allows pulse arrival times to be predicted over long time scales so that small deviations from solar system and astrophysical effects can be determined \citep{Verbiest+2009}. Radio emission beams appear to be locked to the crust of the neutron star and single pulses have phases that vary with respect to a fiducial phase that is also locked to the crust \citep{Kramer1998,cs2010}. Averages of $N_p$ single pulses at a specific frequency  converge to a stable pulse shape with fractional deviations $\sim 1 / \sqrt{N_p}$, as expected for pulse fluctuations that are largely statistically independent \citep[e.g.,][]{Dolch+2014}. While average pulse shapes do vary with frequency \citep{Kramer+1998}, the pulse shapes of radio pulsars, including those objects having two or more stable shapes associated with metastable state of the magnetosphere (i.e., the shapes do not show evolution in time), are stable and show no secular evolution except for a few pulsars in NS-NS binaries where geodetic precession alters the orientation of the beam \citep{Perera+2010} and in the Crab pulsar in which larges changes in pulse shape are seen over a few decades \citep{Lyne+2013}. Magnetars also show secular changes in pulse shapes \citep[e.g.,][]{YanZ+2015}.


Intrinsic variations in pulses appear to have stationary statistics \citep{Liu+2011,Liu+2012} in the same way that the average profile formed by averaging a large number of single pulses converges to a shape that appears to be epoch independent \citep[see][]{Craft1970,brc1975,pw1992,Hassall+2012,Pilia+2015}. Consequently, pulse-to-pulse variations can be characterized for each pulsar and can be incorporated into timing studies  that require a noise model, such as GW detection. Within a Bayesian framework, the average pulse profile and the pulse variations comprise some of the prior information that underlie modeling of pulsar orbits and GW detection \citep{vH+2009,Lentati+2014}.

In this paper, we focus on timescales smaller than one day and as short as a single spin period. Longer time spans require consideration of other phenomena, including pulsar spin variations and changes in the free-electron content along the line of sight. Intrinsic pulse variations comprise only one contribution to the arrival time variance on short timescales. A second contribution is  the template-fitting error due to additive noise in the measured pulse shape which therefore, unlike single pulse variations, depends on the signal-to-noise ratio of the average pulse \citep{cs2010}. A third contribution is due to  changes in the interstellar impulse response from multipath scattering, which depends strongly on radio frequency \citep{cwd+1990}. The measured impulse response (or pulse broadening function, PBF) at a given time is caused by diffractive interstellar scattering/scintillation (DISS) and it varies as the finite number of constructive intensity maxima (`scintles') appearing in the measurement bandwidth changes.  These white-noise contributions to arrival-time errors are referred to as pulse jitter, template-fitting errors, and scintillation noise, respectively. They have distinct correlations with time and frequency that can used to separate them empirically.

In \S\ref{sec:model}, we describe the white-noise model. In \S\ref{sec:observations}, we briefly describe observations from the North American Nanohertz Observatory for Gravitational Waves (NANOGrav) and the data sets used in our analysis. We discuss the analysis  of individual objects in \S\ref{sec:single_analysis}, discuss the collective results in \S\ref{sec:analysis_summary}, and analyze pulse jitter statistics in MSPs in \S\ref{sec:jitter_summary}.  In \S\ref{sec:implications} we compare our results with the parameterized Bayesian noise analysis reported in \citet{NG9yr} and discuss the implications for pulsar timing array (PTA) optimization. We summarize our conclusions in \S\ref{sec:conclusions}.

\section{Model for Short-term Timing Variance}
\label{sec:model}

We characterize the three white-noise contributions through appropriate analysis of short ($\sim 30$~min) timing observations. Typical observing epochs are separated by several days or weeks, over which time each of the three contributions is uncorrelated, thus appearing as a white-noise perturbation of arrival times, $\Delta t(\nu,t)$. The total combined variance of the residuals\footnote{Residuals $\equiv$ (data$-$model), as discussed in \S\ref{sec:observations}. For the white-noise errors we consider, there is little difference between the pre-and-post-fit variance. The differences are discussed in \S\ref{sec:single_analysis} and Appendix~\ref{appendix:deviations}.} on short timescales is
\be
\sigma_\R^2 = \sigma_\SN^2 + \sigma_\J^2 + \sigma_\DISS^2,
\label{eq:white_noise_model}
\ee
where $\sigma_\SN$ is the template-fitting error from a finite pulse signal-to-noise ratio (S/N) primarily due to radiometer noise, $\sigma_\J$ is the error due to pulse phase and amplitude jitter, and $\sigma_\DISS$ is due to scintillation noise.  Spin noise, measurable over roughly yearly timescales, is negligible over a single epoch, as are changes in dispersion measure ($\DM = \int dl\,n_e$, the integral of the electron density over the line of sight) and in the mean shape of the PBF (see Appendix~\ref{appendix:deviations} for more details). For most objects we find $\sigma_\SN > \sigma_\J  \gg \sigma_\DISS$, while a few have $\sigma_\J \gtrsim \sigma_\SN$ at some epochs of high S/N from periods of strong scintillation. Several objects show $\sigma_\DISS$ as the dominant timing error at particular radio frequencies (see \S\ref{sec:analysis_summary}).

In the following, we will consider the pulse shape model and individually discuss the TOA errors resulting from template fitting of finite S/N pulses, jitter, and scattering.

\subsection{Pulse Shapes}

Radio pulses are subject to a variety of perturbations as they travel between the pulsar and the Earth. To model the changes in pulse shape and intensity, we will assume that all chromatic delays have been perfectly removed or are negligible over each narrowband channel. These include the dispersive delay from DM, scattering, and frequency-dependent pulse profile evolution. We also assume that the signal polarization has been calibrated perfectly.

Under these assumptions, we model pulse shapes $I(\phi,\nu,t)$ as a function of phase $\phi$ obtained in short integrations longer than the pulse period, centered on time $t$ and in a sub-band centered on frequency $\nu$. The dominant remaining effect from scattering is the DISS intensity modulation associated with a small number of scintles in a time-frequency resolution cell. Refractive interstellar scintillation (RISS) will also modulate the signal strength but typically varies more slowly than DISS and is broadband \citep[though still chromatic;][]{Stinebring+2000}. It is assumed in the following discussion that we can resolve relevant pulse structure and scintillation fluctuations though in reality observing practices may not always allow for scintles to be fully resolved for a given pulsar.  We also include a telescope bandpass function $H_{\mathrm{tel}}(\nu)$ that lumps together all frequency-dependent gains from the feed antenna to the output of the digital filterbank channel. The pulse shape model is then
\ba
I(\phi,\nu,t) &=& H_{\mathrm{tel}}(\nu)\left\{g_\RISS(\nu,t) g_\DISS(\nu,t) \times \right. \nonumber \\
&& \left. \left[S_i(\nu) p_i(\phi,\nu,t) \ast h_\PBF(\phi,\nu,t)\right] \right. \nonumber \\
&& \left. + n(\phi,\nu,t) \right\}
\label{eq:pulse_model}
\ea
where $g_\RISS$ is the RISS modulation, $g_\DISS$ is the DISS modulation, $S_i$ is the intrinsic spectrum of the pulsar, $p_i$ is the intrinsic pulse shape normalized to unit area, $h_\PBF$ is the pulse broadening impulse response function, and $n$ is additive radiometer noise. The intrinsic pulse shape is stochastic and includes contributions from phase and amplitude jitter. We assume that the time-averaged intrinsic pulse shape, $\langle p_i(\phi,\nu,t)\rangle_t$, converges to a pulse template, $U(\phi,\nu)$, that is stable over long timescales. The template shape evolves as a slow function of frequency and the shape of each individual pulse is as well.

\subsection{Template-Fitting Errors}

Template matching yields an RMS error in the TOAs that depends on the $\SN$ of the pulse. We assume for now that the data profile is a scaled and shifted version of the template with additive noise, the condition for matched filtering to yield the minimum possible TOA error \citep{Turin1960,Taylor1992}. This assumption breaks down when considering pulse phase jitter and the finite scintle effect, which change the profile dynamically and are discussed in the following subsections. Let $U(\phi)$ be the pulse template as a function of pulse phase $\phi$ normalized to unit amplitude, where we have dropped the explicit frequency dependence. The measured pulse intensity $I(\phi)$ at any epoch is then modeled as 
\be
I(\phi) = S\sigma_{\mathrm{n}} U(\phi-\phi_0) + n(\phi),
\label{eq:template_matching}
\ee
where $S$ is the signal-to-noise ratio of the pulse profile (peak to off-pulse RMS, written this way for clarity as a variable in equations), $n(\phi)$ is additive noise with RMS amplitude $\sigma_{\mathrm{n}}$, and $\phi_0$ is the TOA. The TOA can be determined either through a cross-correlation analysis with proper interpolation of the cross correlation function to find the maximum or by least-squares fitting of the model template to the data. Mathematically, the two approaches are identical. The peak of the cross-correlation function (CCF) of the template and pulse profile has a S/N related to $S$ as (J.~M.~Cordes et al.\ in preparation)
\be
S_{\CCF} = S\left[\sum_{i=0}^{\Nphi-1} U^2(\phi_i) \right]^{1/2}
\label{eq:S_TS}
\ee
and is larger by a factor equal to the square root of the effective number of samples across the pulse if $n(\phi)$ is uncorrelated between phase bins. Template matching will fail when $S_{\CCF} \lesssim 1$.

For a pulse template with $\Nphi$ phase bins, the template-fitting error is \citep{cs2010}
\be
\sigma_\SN = \frac{\Weff}{S \sqrt{\Nphi}},
\label{eq:radiometer_noise}
\ee
where $\Weff$ is an effective width\footnote{This is a different definition than given in \citet{cs2010} although the RMS error expressions are the same.} given by 
\be
\Weff = \frac{P}{\Nphi^{1/2}\left[\mathlarger{\sum}\limits_{i=1}^{\Nphi-1} \left[U(\phi_i) - U(\phi_{i-1})\right]^2\right]^{1/2}}
\label{eq:Weff}
\ee
for a pulsar with period $P$. We note that for Eq.~\ref{eq:radiometer_noise}, if profiles are smoothed by $n_s$ samples to increase $S \propto n_S^{1/2}$, the effective number of phase bins $\Nphi \propto n_s^{-1}$, leaving the product $\Nphi^{1/2}S$ invariant. The effective width  is useful because it is unique to each pulsar-frequency combination and does not depend on any observational parameters. Therefore, it can be calculated using data obtained from one receiver-backend system and then the TOA error can be calculated for any value of $\SN$ and number of phase bins.  Any instrumental change, such as a change in $H_{\rm tel}(\nu)$ over time, that alters the pulse shape will have to be taken into account, however. The expression for $\sigma_\SN$ yields the same value as the frequency-domain expression given by \citet{Taylor1992}.

\subsubsection{The Role of DISS}

The finite S/N causes the TOA to have a Gaussian error PDF under the assumption of the central limit theorem, $f_{\Delta t}(\Delta t | S) = \mathcal{N}(0,\sigma_{\SN}^2)$. DISS  causes the $\SN$ of the pulse to be modulated by a scintillation ``gain'', $g$. The gains have an exponential PDF
$f_g(g) = \exp(-g)\Theta(g)$ where $\Theta(g)$ is the Heaviside step function
\citep[see Appendix B of][]{cc1997}. Multiple scintillation maxima in the time-frequency plane will alter the PDF, which, given $\niss$ scintles, is 
\be
f_g(g \vert \niss) = \frac{(g\niss)^{\niss}}{g\Gamma(\niss)} e^{-g\niss}\Theta(g),
\label{eq:scintillation_pdf}
\ee
where  $\Gamma$ is the gamma function. When pulse shapes and TOAs are calculated, typically $\niss \gtrsim 1$ scintles are averaged over the bandwidth and integration time, decreasing the variations in the scintillation gains.   

We can transform the PDF of gains to the PDF of the observable pulse S/Ns with a change of variable to $g = S/S_0$, where $S_0$ is the mean S/N. The PDF is written as
\be
f_S(S \vert \niss) = \frac{(S\niss/S_0)^{\niss}}{S\Gamma(\niss)} e^{-S\niss/S_0}\Theta(S).
\label{eq:sn_pdf}
\ee
As $\niss\to\infty$, $f_S(S\vert \niss) \to \delta(S-S_0)$, and the pulse S/N will be constant.

The PDF of the TOA errors is
\ba
f_{\Dt}(\Dt \vert \niss) &=& \frac{1}{\sigma_{S_0} \sqrt{2\pi}} \left(\sqrt{2}\niss \frac{\sigma_{S_0}}{\left|\Dt\right|}\right)^{\niss+1} \times \nonumber \\
&& H_{-\left(\niss+1\right)}\left(\frac{\niss \sigma_{S_0}}{\sqrt{2}\left|\Dt\right|}\right)
\label{eq:scintillation_toas_pdf}
\ea
where $\sigma_{S_0}$ is the RMS from template-fitting errors when no scintillation occurs ($S$ is constant) and $H_n(x)$ is a Hermite polynomial of order $n$. See Appendix~\ref{appendix:pdfs} for more details.  In general, the distribution of measured $\SN$, $f_S(S)$, will be a convolution of several distributions, including the distribution of $\SN$ intrinsic to the pulsar $f_{S_{\rm int}}(S)$, the DISS modulation $f_{S_{\DISS}}(S)$, and the RISS modulation $f_{S_{\RISS}}(S)$, which will also affect the distribution of TOA errors.

\subsubsection{Example of a Single-Component Gaussian Pulse}

For a Gaussian pulse having width W (FWHM), the effective width (using Eq.~\ref{eq:Weff}) is
\be
\Weff = \frac{(WP)^{1/2}}{(2\pi\ln 2)^{1/4}}
\label{eq:Weff_Gaussian}
\ee
For this case, the effective width is proportional to  the geometric mean of the period and actual pulse width. The TOA error is
\be
\sigma_{\SN}= \frac{\left(WP\right)^{1/2}}{\left(2\pi \ln 2 \right)^{1/4} \Nphi^{1/2} S} = \frac{W}{2\left(\ln 2\right)^{1/2} S_{\CCF}},
\label{eq:sigmaSN_Gaussian}
\ee
where we have used Eq.~\ref{eq:S_TS} to calculate
\ba
S_{\CCF} &=& \frac{S}{2}\left(\frac{2\pi}{\ln 2}\right)^{1/4} \left(\frac{W \Nphi}{P}\right)^{1/2}\nonumber\\
&\approx& 5.55 \left[\left(\frac{W/P}{0.02}\right) \left(\frac{\Nphi}{2048}\right)\right]^{1/2} S.
\ea
The quantity $W/P$ represents the fiducial duty cycle for an MSP. $S_{\CCF}$ must be of order unity or larger for template matching to fit appropriately. 


\subsection{Single Pulse Amplitude and Phase Variations (``Jitter'')}

Single pulses of both canonical pulsars and MSPs have been shown to have stochastic amplitude and phase variations \citep{cd1985,cwd+1990,Liu+2012,sc2012,sod+2014,Dolch+2014}. When averaged over $\Np$ pulses to form a pulse profile, pulse jitter causes the underlying pulse shape to differ from that of the template, causing an error that is qualitatively different from additive noise. The jitter TOA error is independent of $\SN$. We define a dimensionless parameter $\kJ \equiv \sigma_{\J,1}/P$ as the ratio of RMS phase variation of individual pulses $\sigma_{\J,1} = \sigma_\J \sqrt{\Np}$ (in time units) to the period $P$ of the pulsar. \citet{cd1985} and \citet{cs2010} define a jitter parameter $\fJ = \sigma_{\J,1} / \sigma_U$, where $\sigma_U$ is the equivalent RMS width of the template. Since pulse profiles often display multiple components with potentially different jitter statistics, using $\kJ$ to compare the intrinsic jitter between pulsars is less dependent on the properties of the different components.

We note that single-component pulses that show phase variations only will have an RMS jitter but those that show amplitude variations only will not display jitter. However, for pulses with multiple components, amplitude variations without phase variations will yield an RMS jitter but only if the components overlap in pulse phase. An in-depth analysis on the role of multiple components in jitter will be presented in J.~M.~Cordes et al.\ (in preparation). As an example, we consider a single component, Gaussian-shaped pulse with both a Gaussian phase jitter PDF with dimensionless phase variations $\kJc$ and amplitude variations with a modulation index $\mIc$ (defined as RMS intensity divided by mean pulse amplitude). We use the subscript `c' to explicitly denote that the parameters describe the single component, whereas the parameter $\kJ$ is defined as the overall timing variation of the pulse. The TOA error is then \citep[modified from the form in][]{cs2010}
\be
\sigma_\J = \frac{\kJ P}{\sqrt{\Np}} = \kJc P \left(\frac{1+\mIc^2}{\Np}\right)^{1/2}.
\label{eq:jitter_noise}
\ee

Comparing the TOA errors from additive noise and jitter in Eqs.~\ref{eq:radiometer_noise} and \ref{eq:jitter_noise}, we can define a transition S/N at which the two contributions are equal, $\sigma_\SN = \sigma_\J$. The single-pulse S/N implied by a profile calculated from $\Np$ pulses, assuming statistical independence of jitter between pulses, is $S_1 = \Np^{-1/2} S$. For a Gaussian-shaped pulse, we find the single-pulse transition S/N, by setting Eqs.~\ref{eq:sigmaSN_Gaussian} and \ref{eq:jitter_noise} equal when $\Np = 1$, to be
\ba
S_{1,\mathrm{trans}} & = & \kJc^{-1} \left(\frac{W}{P}\right)^{1/2} \left(2 \pi \ln 2\right)^{-1/4} \left[\Nphi\left(1+\mIc^2 \right)\right]^{-1/2} \nonumber \\
&  \approx & 0.216 \left(\frac{\kJc}{0.007}\right)^{-1} \left(\frac{W/P}{0.02}\right)^{1/2}\nonumber\\
& & \times \left(\frac{\Nphi}{2048}\right)^{-1/2}\left(\frac{1+\mIc^2}{2}\right)^{-1/2}
\ea
and the corresponding S/N of the CCF is 
\ba
S_{\CCF1,\mathrm{trans}} &\approx &1.20 \left(\frac{\kJc}{0.007}\right)^{-1} \nonumber\\
& & \times \left(\frac{W/P}{0.02}\right) \left(\frac{1+\mIc^2}{2}\right)^{-1/2},
\label{eq:transition_SN}
\ea
where we set the fiducial $\kJ = \kJc(1+\mIc^2)^{1/2} = 0.01$ based on our analysis in \S\ref{sec:jitter_summary}. When the single-pulse cross-correlation $\SN$ is greater than about unity, the jitter error becomes larger than the template-fitting error.

The same pulsar-intrinsic effects that cause frequency-dependent template evolution will cause jitter to be a slow function of frequency as well. Over an observing band, we might approximate jitter as being frequency-independent (see \citealt{sod+2014} for evidence of decorrelation over widely-separated frequencies) but frequency-dependence of the pulse template can be measurable \citep{pdr2014,Dolch+2014}. We therefore note that jitter will be strongly correlated in frequency but not in time. DISS has a correlation bandwidth and timescale that can vary widely from pulsar to pulsar and between epochs for the same pulsar. Template-fitting errors are uncorrelated between time samples and frequency sub-bands.

\subsection{Scintillation Timing Noise: Finite Scintle Effect}

The time-frequency plane is made up of independent intensity fluctuations called scintles that are 100\% modulated and have characteristic time and frequency scales $\Dtd$ and $\Dnud$, respectively. The scintillation structure is related to the temporal broadening of pulses, resulting in a time delay \citep{cwd+1990,cs2010}. Since a finite number of scintles will occupy the time-frequency plane, the instantaneous PBF will be different from the ensemble average shape. This produces an error that is statistically independent between two epochs and is therefore white noise in time.

The number of scintles for an observation of duration $T$ and bandwidth $B$ is approximately
\be
\niss \approx \left(1+\eta_t \frac{T}{\Dtd}\right)\left(1+\eta_\nu \frac{B}{\Dnud}\right).
\label{eq:N_DISS}
\ee
The filling factors $\eta_t$,$\eta_\nu$ are less than unity and are in the range of 0.1 to 0.3 \citep{cs2010,Levin+2016}, depending on the definitions of the characteristic timescale and bandwidth.

When $\niss$ is large, the TOA error is
\be
\sigma_\DISS \approx \frac{\taud}{\sqrt{\niss}}
\label{eq:diss_noise}
\ee
where $\taud = C_1 / (2\pi\Dnud)$ is the scattering timescale with $C_1$ a coefficient of order unity. For a thin scattering screen with uniscale irregularities, $C_1 = 1$ but for a Kolmogorov screen, $C_1 = 0.96$. For uniform, thick media, $C_1 = 1.53$ and 1.16, respectively, for uniscale and Kolmogorov media \citep{cr1998}. When there is only one scintle or a partial scintle across the band, the TOA error is approximately $\taud$, or some fraction of it.

\section{Observational Data}
\label{sec:observations}

\subsection{NANOGrav Timing Observations}

\begin{figure*}[t!]
\epsscale{1.2} 
\begin{center}
\plotone{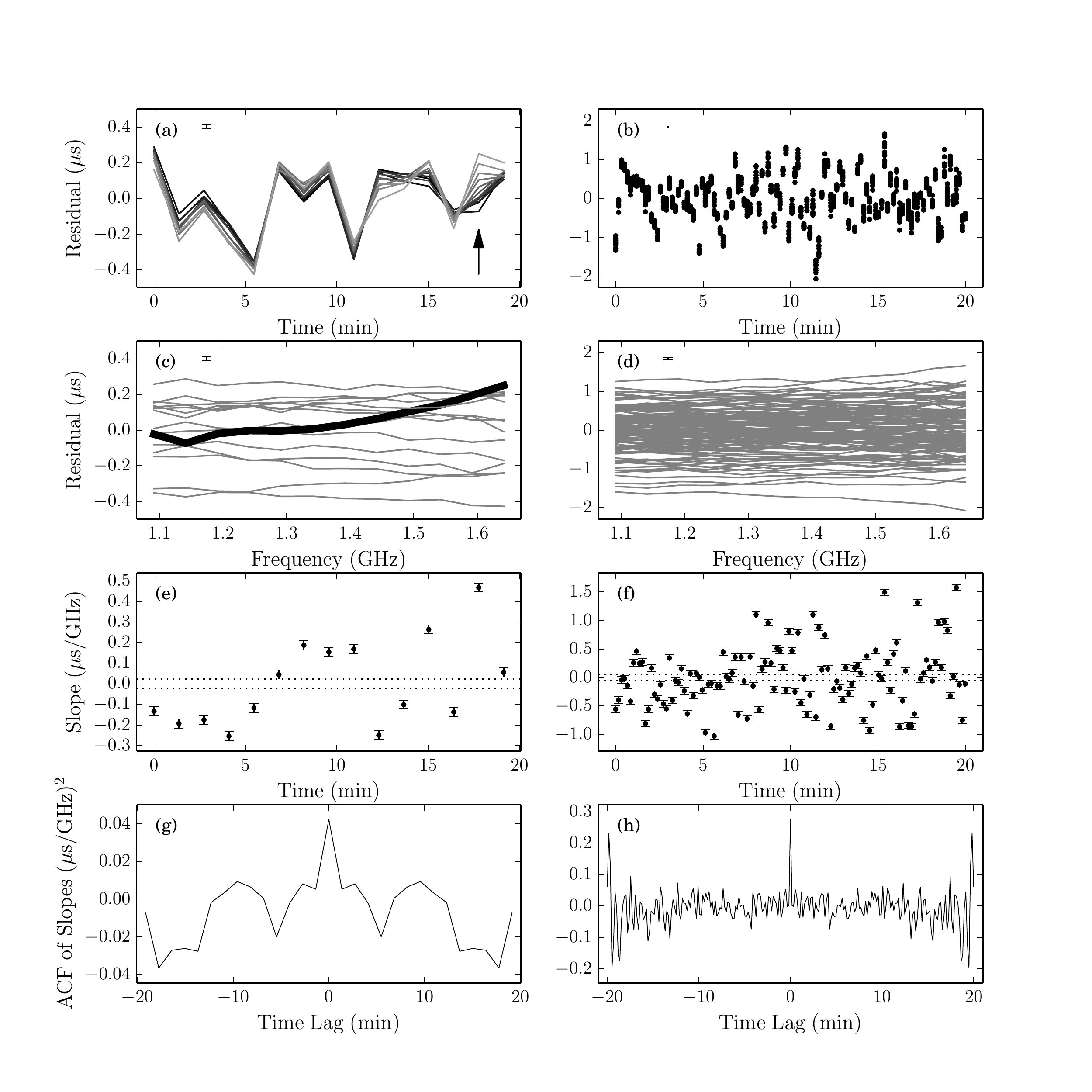}
  \caption{\footnotesize Analysis of jitter in residuals for the highest $\SN$ epoch for PSR J1713+0747. The panels on the left side show the analysis for the $\sim 80$~s subintegration data while the panels on the right side are for the $\sim 10$~s subintegration data. Panel (a): Low-time-resolution residuals as a function of time. Each frequency channel is shaded differently, with darker lines indicating lower frequencies. The arrow indicates the subintegration with the greatest change in residual versus frequency. Typical TOA errors are shown in the top left of panels (a)-(d). Panel (b): High-time-resolution residuals as a function of time, where we have plotted residuals as points for clarity. Panels (c),(d): Residuals as a function of frequency, where each line represents one subintegration. The thick, black line in panel (c) corresponds to the subintegration highlighted with the arrow in panel (a). Panels (e),(f): Slopes of fitted lines to the residuals versus frequencies for each subintegration. The horizontal, dotted lines indicate the median fitting error. Panels (g),(h): Autocorrelation functions (ACFs) of the time series in panels (e) and (f), respectively.}
\label{fig:highsn_epoch}
\end{center}
\end{figure*}

\begin{deluxetable*}{llcl}
\centering
\tablecolumns{4}
\tablecaption{Errors in Initial Timing Model}
\tablehead{
\colhead{Effect} & \colhead{Typical $\Delta t$} & \colhead{Appendix Section} & \colhead{Comments}\\
}
\startdata
Pulse profile smearing & & & \\
\hline
spin period error & $\lesssim$10~ps  & A.1.1 & systematic\\
binary parameter errors & $\lesssim$10~ns  & A.1.2 & systematic\\
DM variations & $\lesssim$400~ns & A.1.3 & stochastic\\
polarization calibration gain errors & $\lesssim$1~$\mu$s & A.1.4 & stochastic\\
\hline
Deviations from the polynomial fit & & & \\
\hline
binary orbit parameter errors & $\lesssim$10~ps & A.2.1 & systematic\\
ionospheric DM variations & $\lesssim$1~ns& A.2.2 & stochastic\\
cross-coupling errors & ? & A.2.3 & systematic, highly pulsar-dependent\\
rotation measure (RM) variations & $\lesssim$1~ps & A.2.4 & stochastic\\
spin noise & $\lesssim$0.1~fs & A.2.5 & stochastic\\
stochastic GW background & $\lesssim$1~fs & A.2.6 & stochastic
\enddata
\label{table:fitconsiderations}
\end{deluxetable*}

We used pulse profile data from the NANOGrav nine-year data set  described in \citet[][hereafter NG9]{NG9yr} for our analysis. NG9 contains multi-frequency pulse profiles of thirty-seven MSPs observed at the Green Bank Telescope (GBT) and Arecibo Observatory (AO). Two generations of backends were used, the GASP/ASP backend earlier, processing up to 64 MHz \citep{Demorest2007,Demorest+2013}, and the GUPPI/PUPPI backends later, processing 100, 200, or 800 MHz of bandwidth \citep{drd+2008,fdr2010}. The larger bandwidth of GUPPI and PUPPI yields an increase in $\SN$ from  increased averaging of radiometer noise combined with a higher probability for large scintillation maxima  \citep{pdr2014}. Because we wish to maintain homogeneity of the inferred parameters of our pulsars (e.g., consistent scintillation statistics), we analyze pulses observed with GUPPI/PUPPI only.

Each pulsar was observed at each epoch with at least two receivers. At GBT, the 820 and 1400~MHz bands were used, and at AO, the 430 and 1400~MHz or 1400 and 2300~MHz bands were used. PSRs B1937+21 and J1713+0747 were observed at both AO and GBT and we analyze both observatories' data sets independently to check for consistency across varying S/Ns. In addition, PSR J2317+1439 contained data from the 327~MHz band in addition to the 430 and 1400~MHz bands. We also used processed 430~MHz data available for PSRs B1937+21 and J2017+0603 though they were not included in NG9.

Pulse profiles were computed in real time by averaging together single pulses according to an initial timing model that includes the pulsar's spin kinematics and the orbital motions of the Earth and, if needed, the pulsar binary orbit. Model parameters were obtained by fitting to earlier observations. Raw data profiles from GUPPI/PUPPI were folded and de-dispersed in $\sim 10$~s and $\sim 15$~s subintegrations at AO and GBT, respectively, and every eight subintegrations were averaged together to reduce data volume through the NG9 pipeline. Some Arecibo 1400~MHz observations were initially recorded in $\sim 1$~s subintegrations to aid in radio frequency interference (RFI) excision and then combined to form the $\sim 10$~s ``raw'' subintegrations. Observations for a given epoch typically spanned about 0.5~hr. All profiles were divided into 2048 phase bins.

\citet{NG9yr} describe the polarization calibration algorithm, as well as the RFI excision methods, for creating calibrated data profiles using the \textsc{psrchive}\footnote{\url{http://psrchive.sourceforge.net}, accessed via scripts available at \url{https://github.com/demorest/nanopipe}} software package \citep{Hotan+2004,vS+2012}. A broadband noise source was locally injected into the two polarization signal paths at each observatory prior to every pulsar observation and is recorded by the backend systems. Both differential gain and phase between the two hands of polarization were calibrated using the correlated noise source observation. The noise source power in each hand of polarization was not assumed to be equal and was measured separately roughly once per month per telescope per frequency by observing the noise source after pointing on and off a bright, unpolarized quasar. After balancing the gains of the two orthogonal polarizations, the intensity profiles were produced by summing the two polarization profiles. Future papers will discuss the complete polarization and flux calibration solutions at AO and GBT. Frequency channels known to consistently contain RFI signals were removed first. If the off-pulse variation in a 20-channel wide frequency window was four times the median variation value, those channels were also removed.

We took the calibrated profiles with $\sim 80$~s (AO) and $\sim 120$~s (GBT) subintegration lengths and average the profiles together into sub-bands of 50~MHz resolution. Frequency-averaging builds $\SN$ for each pulse to avoid mis-estimation of the TOA in the low-$\SN$ limit \citep[see Appendix B of][]{NG9yr}. We note that frequency-dependent profile shape changes  across the entire observing band can be significant for some sources over the full band \citep[e.g., see][]{pdr2014} but are small over a 50~MHz channel.

We implemented a Fourier-domain TOA estimation algorithm \citep{Taylor1992} that determines the amplitude $S\sigma_n$, the TOA, and template-fitting uncertainty of an intensity profile $I(\phi,\nu,t)$. Template shapes $U(\phi)$ are determined from de-noised average profiles, smoothed by thresholding the coefficients of a wavelet decomposition of the pulse shape. One template is generated from all data for each pulsar, backend, and frequency band combination\footnote{Templates are available in the NG9 data release at \url{https://data.nanograv.org}}. Timing offsets from profile frequency evolution are not accounted for here but will be accounted for in the analysis in the following section. We determined the off-pulse window for each pulse template used to measure $\sigma_n$ by finding the rolling eighth (256 out of 2048 phase bins) of phase that has the smallest integrated intensity. The pulse baseline is defined as the mean of the off-pulse region and the noise $\sigma_n$ is the RMS amplitude of the region. Once we knew the best-fit amplitude and RMS noise, we then calculated the associated $\SN$ for each pulse. Our code is freely available in the PyPulse software package\footnote{https://github.com/mtlam/PyPulse}.

\subsection{Scintillation Parameters}

Scintillation bandwidths and timescales were taken or estimated (using the scaling relations as a function of observing frequency in \citealt{NE2001}) from \cite{Keith+2013} and \cite{Levin+2016} and references therein. We used these measurements to derive values of $\sigma_\DISS$ given by Eq.~\ref{eq:diss_noise} assuming $\eta_\nu = \eta_t = 0.2$, $C_1 = 1$ \citep{lr1999,cs2010,Levin+2016}, and integration time/bandwidth values equal to that of the profiles from each telescope. When scintillation parameters were not available, we estimated all other values using the NE2001 electron density model \citep{NE2001}.


\section{Single Pulsar Analysis} 
\label{sec:single_analysis}

We are interested in quantifying noise on intraday timescales. We therefore independently analyze individual NANOGrav observations, typically of duration 30~min or less. During an observation, the incoming data were folded using a pre-computed model pulsar ephemeris. We assumed that this ephemeris is sufficiently accurate that there is very little drift in pulse arrival times over an observation. We calculated pulse phases within an observation, ``initial timing residuals'' $\delta t(\nu,t)$, using the Fourier-domain estimation algorithm of \citet{Taylor1992}. We assumed that the initial timing model used for folding will yield polynomial expansions of phase and spin period that represent the state of the Earth-pulsar line of sight at a given epoch to high accuracy. We also assumed that the initial timing model is accurate such that pulse smearing will be negligible for our subintegration lengths. We then calculated ``short-term'' residuals $\R(\nu,t)$ over a single observation by fitting a polynomial model over all $\delta t(\nu,t)$ observed that includes a constant offset for TOAs from each frequency channel and a parabolic fit in time common to all TOAs. The initial and short-term models can be written as 
\ba
\delta t(\nu,t) & = & K(\nu) + a t + b t^2 + n(\nu,t)\\
\R(\nu,t) & \equiv & \hat{n}(\nu,t) = \delta t(\nu,t) - \left[\hat{K}(\nu) + \hat{a}t + \hat{b}t^2\right].
\label{eq:quadratic_fit}
\ea
Here, $a$ and $b$ are frequency-independent coefficients, $n(\nu,t)$ is additive noise in both time and frequency that includes the three white-noise components in Eq.~\ref{eq:white_noise_model}, and $K(\nu)$ represents a constant offset that varies with frequency, resulting from pulse profile evolution or epoch-dependent dispersion and scattering. Variables with carets denote estimated quantities. Thus, $\R(\nu,t)$ is the estimated additive noise, calculated by subtracting the estimated model parameters from the TOAs. We assumed that subtraction of the offsets removes any frequency-dependence between sub-bands. The variance removed by the fit for $a$, $b$, and $K(\nu)$ to obtain $\R(\nu,t)$ will be small for white-noise components that are uncorrelated in time.

Differences between the initial timing model and the short-term timing model for a given epoch can result from a number of possible effects that we account for with the quadratic fit in Eq.~\ref{eq:quadratic_fit}. Table~\ref{table:fitconsiderations} lists the effects and their approximate amplitudes. We provide details of the estimates in Appendix~\ref{appendix:deviations}.

\subsection{An In-Depth Analysis of Jitter and Frequency-dependent Jitter Evolution in PSR~J1713+0747}

PSR J1713+0747 is not only one of the best-timed pulsars but it is the pulsar with the highest $\SN$ pulses in our entire data set and is thus most sensitive to jitter error. The $\SN$ peaked at $S\approx 2000$ at 1400~MHz for one of two observations on MJD 56380. Figure~\ref{fig:highsn_epoch} shows the residuals of sub-bands for that observation in panel (a); strong correlation between sub-bands is evident and indicative of pulse jitter \citep{cs2010,sod+2014}. Along with the $\sim 80$~s data, we processed the $\sim 10$~s subintegration raw data from this observation to demonstrate the lack of temporal correlation in the residuals, shown in panel (b), as expected when jitter noise becomes dominant. The results in Figure~\ref{fig:highsn_epoch} are presented with the 80~s subintegrations displayed in the left panels and the 10~s subintegrations on the right.

Within each subintegration, we typically saw a monotonic increase or decrease in the residual with frequency, which is most evident in the second-to-last subintegration of the low-time-resolution residuals, highlighted with a black arrow in panel (a). Each line in panels (c) and (d) shows the residuals as a function of frequency for each subintegration. The subintegration highlighted with the arrow in panel (a) is marked with a thick, black line in panel (c) and demonstrates the trend increasing with frequency. We fit the slope of each line and plot the results in panels (e) and (f). The slope values show a deviation much larger than the median of the fitting errors, denoted by the dotted, horizontal lines. While the points in the time series in panel (e) appear correlated, they do not in panel (f), suggesting the slope changes are uncorrelated. The autocorrelation functions (ACFs) of the time series in panels (e) and (f) are shown in panels (g) and (h), respectively, and the flatness at non-zero lags demonstrates that the slope changes are consistent with uncorrelated, white noise in time.


The roughly monotonic slope of the residuals with frequency in each subintegration indicates that there is a systematic variation of the pulse shape for each subintegration (and thus TOAs) versus frequency, indicative of frequency-dependent jitter evolution, which is distinctly different from frequency-dependent pulse profile evolution, though related. These slopes are uncorrelated between subintegrations, indicating that longer averages of larger numbers of pulses will show less variation with frequency. Nonetheless, it is known that the average pulse shape of PSR J1714+0747 varies systematically with frequency \citep{NG9yr,Dolch+2014} and those must reflect the variations occurring on the single-pulse level. For the high-time-resolution data, the RMS slope is $\approx 0.53~\mathrm{\mu s/GHz}$.

\begin{figure*}[t!]
\epsscale{1.0} 
\begin{center}
\plotone{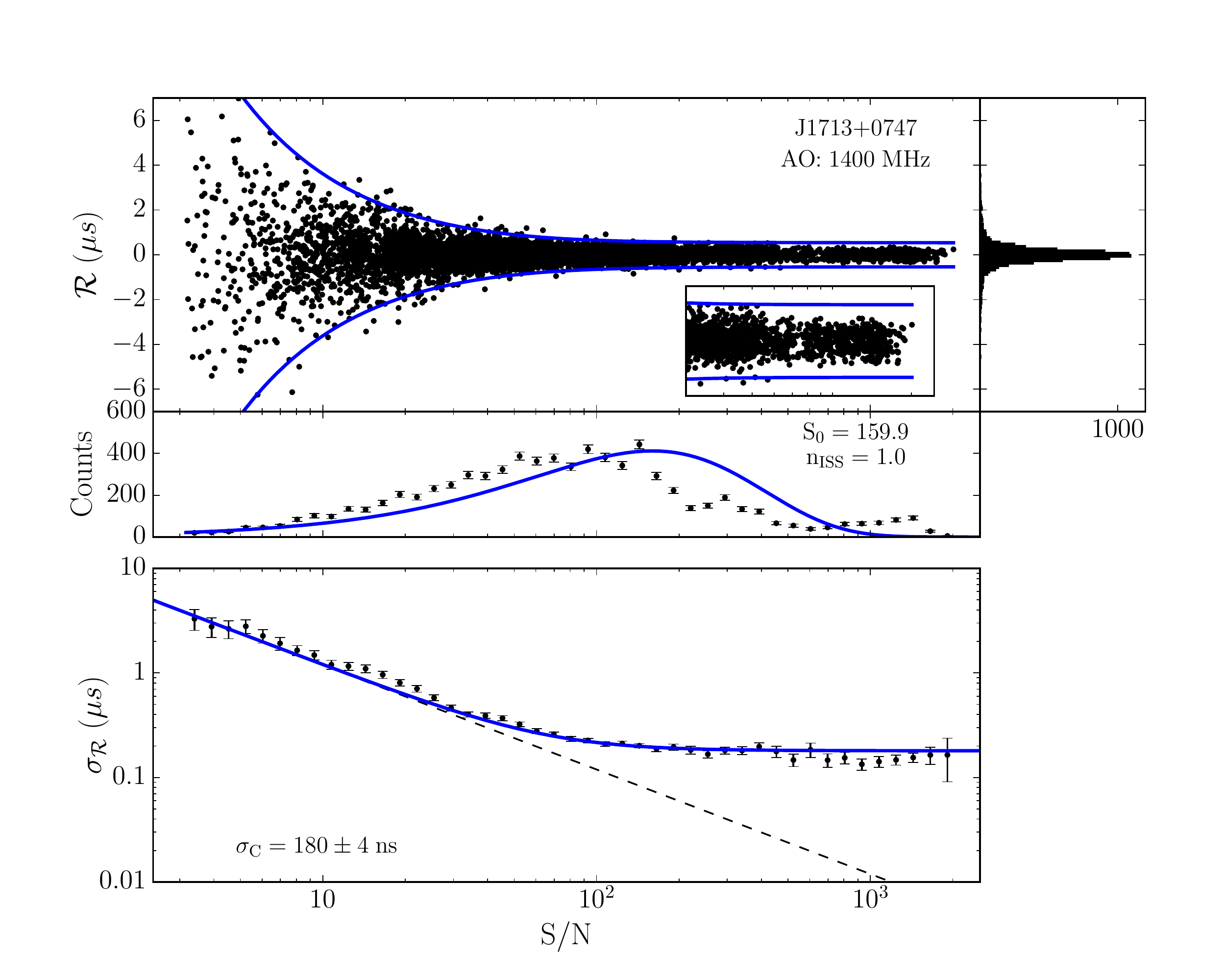}
  \caption{\footnotesize Analysis of residuals for PSR J1713+0747 observed at 1400~MHz with AO, containing the highest $\SN$ residuals in our sample. Top: Residuals $\R$ vs $\SN$. The solid lines (blue) show the $\pm$3$\sigma_\R(S)$ ranges from the maximum likelihood analysis. The inset shows the residuals for S/N greater than 70\% of the maximum. Histograms of $\R$ (right panel) and $\SN$ (middle panel) are shown, with the solid (blue) lines showing the predicted histogram given the most-likely estimates for $S_0$ and $\niss$. The error bars show the standard Poisson uncertainties for each bin only. Bottom: RMS residual $\sigma_\R$ in bins of $\SN$. The dashed line is the predicted TOA template-fitting error (not a fit to the points on the graph) based on the template shape while the solid line shows the estimated $\sigma_\R(S)$ from Eq.~\ref{eq:one-parameter_model} that includes a S/N-independent term.}
\label{fig:J1713+0747_L-wide_PUPPI}
\end{center}
\vspace{-2ex}
\end{figure*}

\begin{figure}[t!]
\hspace{-5ex}
\includegraphics[width=0.525\textwidth]{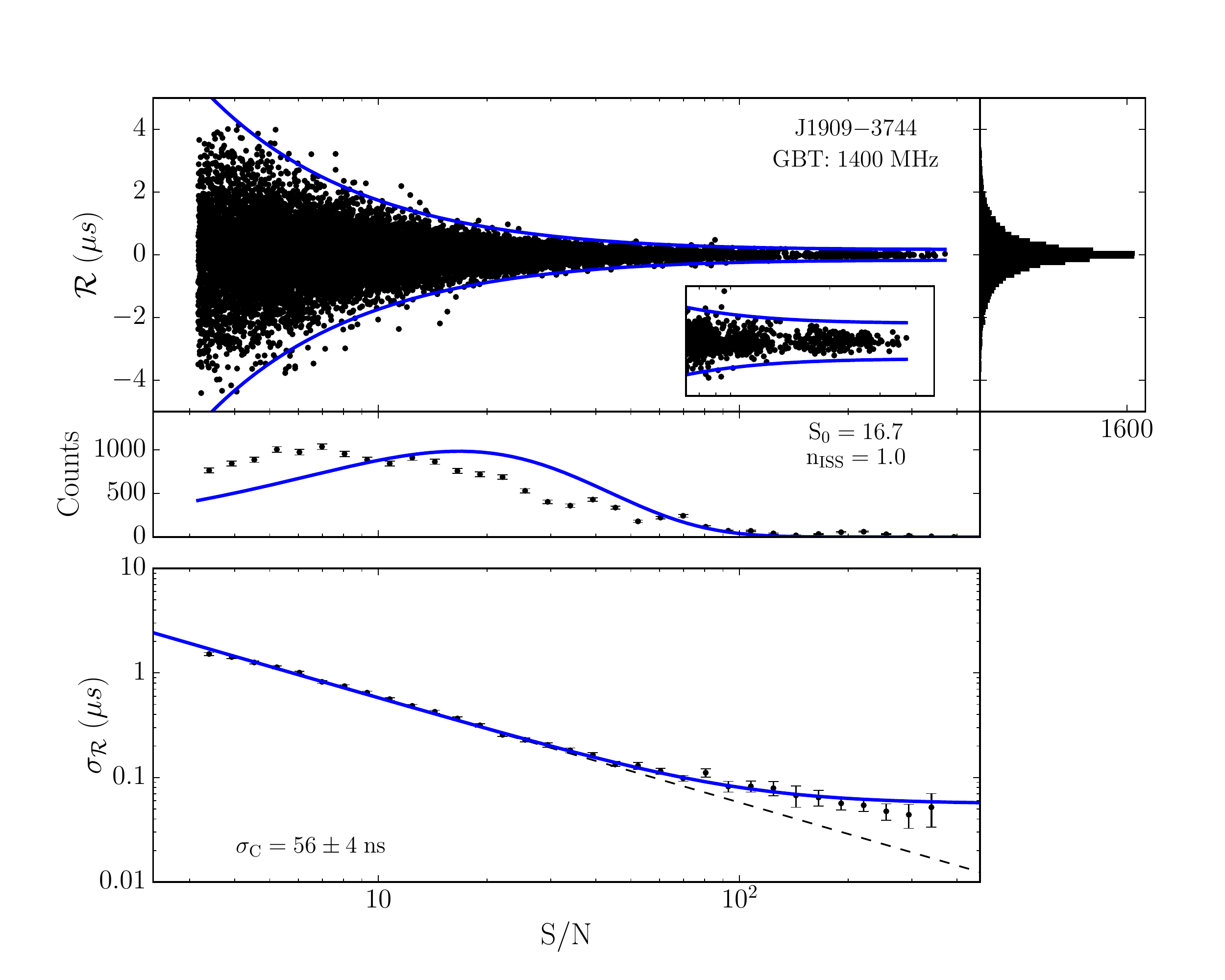}
  \caption{\footnotesize Analysis of residuals for PSR J1909$-$3744 observed at 1400~MHz with GBT. See the Figure \ref{fig:J1713+0747_L-wide_PUPPI} caption for more details.\vspace{3ex}}
\label{fig:J1909-3744_Rcvr1_2_GUPPI}
\end{figure}

\begin{figure}[t!]
\hspace{-5ex}
\includegraphics[width=0.525\textwidth]{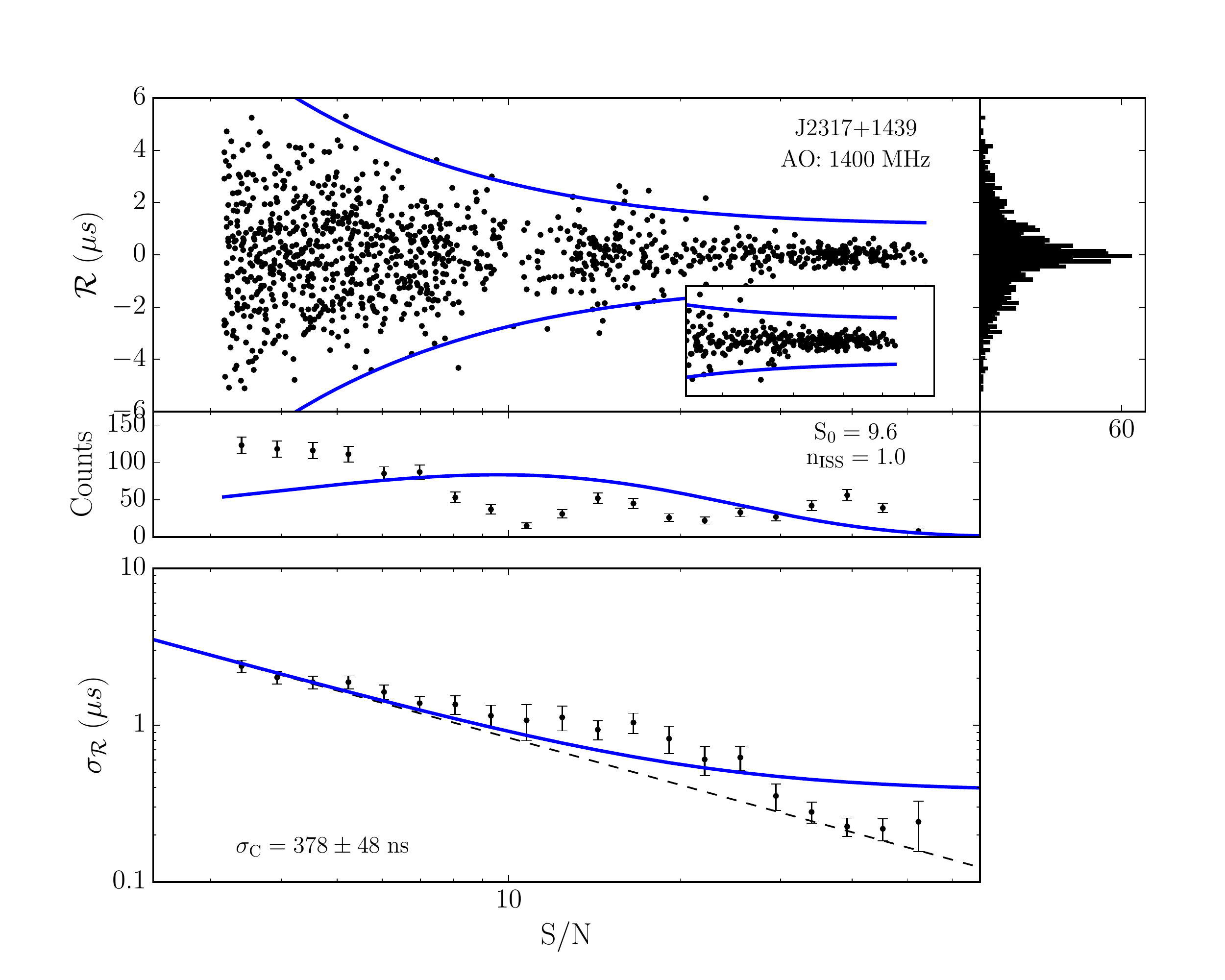}
  \caption{\footnotesize Analysis of residuals for PSR J2317+1439 observed at 1400~MHz with AO. See the Figure \ref{fig:J1713+0747_L-wide_PUPPI} caption for more details. \vspace{2ex}} 
\label{fig:J2317+1439_L-wide_PUPPI}
\end{figure}

\subsection{Distributions of Residuals from Jitter and Scintillation}
\label{sec:distributions}

We modeled the variance in the residuals separately for each pulsar/backend/frequency band combination using Eq.~\ref{eq:white_noise_model}. While all three terms scale as $\Np^{-1}$, only the template-fitting term depends on the $\SN$ of the pulse profile whereas the jitter and the DISS terms do not. Therefore, we used a one-parameter model for the variance as a function of $\SN$,
\be
\sigma_\R^2(S) = \sigma_\SN^2(S) + \sigma_\C^2 = \left(\frac{\Weff}{S\sqrt{\Nphi}}\right)^2 + \sigma_\C^2. 
\label{eq:one-parameter_model}
\ee
where $\sigma_\C^2 = \sigma_\J^2 + \sigma_\DISS^2$, as implied by Eq.~\ref{eq:white_noise_model}, is the variance that is constant in S/N. At high $\SN$, $\sigma_\SN \to 0$ and $\sigma_\C^2$ becomes the dominant term. We took the scintillation parameters to be constant for all epochs so that $\sigma_{\DISS}$ is fixed, though measurements of these parameters indicate small variations (factor of $\lesssim 2$) over many years, with some pulsar showing larger fluctuations \citep[e.g.,][]{Coles+2015}.

The observed $\SN$ PDF depends on the intrinsic pulse amplitude distribution, on modulations from DISS and RISS, and on variations of the system equivalent flux density (SEFD) of the receiver. We assumed that the average {\it intrinsic} flux density of the pulsar and SEFD were constant over all times. Therefore, the mean \SN, $S_0$, is constant for our many-period pulse averages (large $\Np$), i.e., $f_{S_0}(S) = \delta(S-S_0)$, assuming that changes in the $\SN$ are due solely to modulation from DISS. RISS has been shown to change the observed flux density by a factor of $\lesssim 2$ on the timescale of 10s of days \citep{Stinebring+2000}. Since we observed $\SN$ variations spanning over an order of magnitude from the mean in some cases, we ignored the contribution to the $\SN$ PDF from RISS.

We assumed that residuals at a given $\SN$ follow a Gaussian distribution
\be
f_{\R|S}(\R|S,\sigma_\C) = \frac{1}{\sqrt{2\pi\sigma_\R^2}} e^{-\R^2 / (2\sigma_\R^2)},
\label{eq:fR}
\ee
where again $\sigma_\R$ is a function of both $S$ and $\sigma_\C$ (Eq.~\ref{eq:one-parameter_model}). The normality assumption is a good approximation due to the fact that while residuals must lie within one cycle of pulse phase, $|\R|\lesssim 0.001 P$ and deviation from a Gaussian distribution is negligible. We removed all residuals with $S < 10^{0.5}$ ($\approx 3$) to avoid contamination by low-significance noise being fit by the template matched filtering \citep[see Appendix B of][]{NG9yr}, which excluded five pulsar/backend/frequency band residual sets and two pulsars from our analysis entirely. We excised evident RFI beyond the methods described in \S\ref{sec:observations} by inspection of the residuals and the corresponding pulse profiles.

We performed a maximum likelihood (ML) analysis over the residuals $\{S_i,R_i\}$ given the three parameters $S_0$, $\niss$, and $\sigma_\C$. To include our cut in $\SN$, we included a parameter $\Smin$ and determined the factor that properly normalizes the distribution in $S$. The normalized distribution is 
\ba
f_S(S|S_0,\niss,\Smin) & = &f_s(S|S_0,\niss) \Theta(S-\Smin) \times \nonumber\\
& & \frac{\Gamma(\niss)}{\Gamma(\niss,\niss \Smin/S_0)},
\label{eq:scintillation_pdf_smin}
\ea
where $\Gamma(\alpha,x)$ is the incomplete Gamma function and $\Gamma(\alpha,0) = \Gamma(\alpha)$ \citep[see Eqs.~3.381.3-4 of][for the relevant integrals]{GR}.

\begin{figure}[t!]
\hspace{-5ex}
\includegraphics[width=0.525\textwidth]{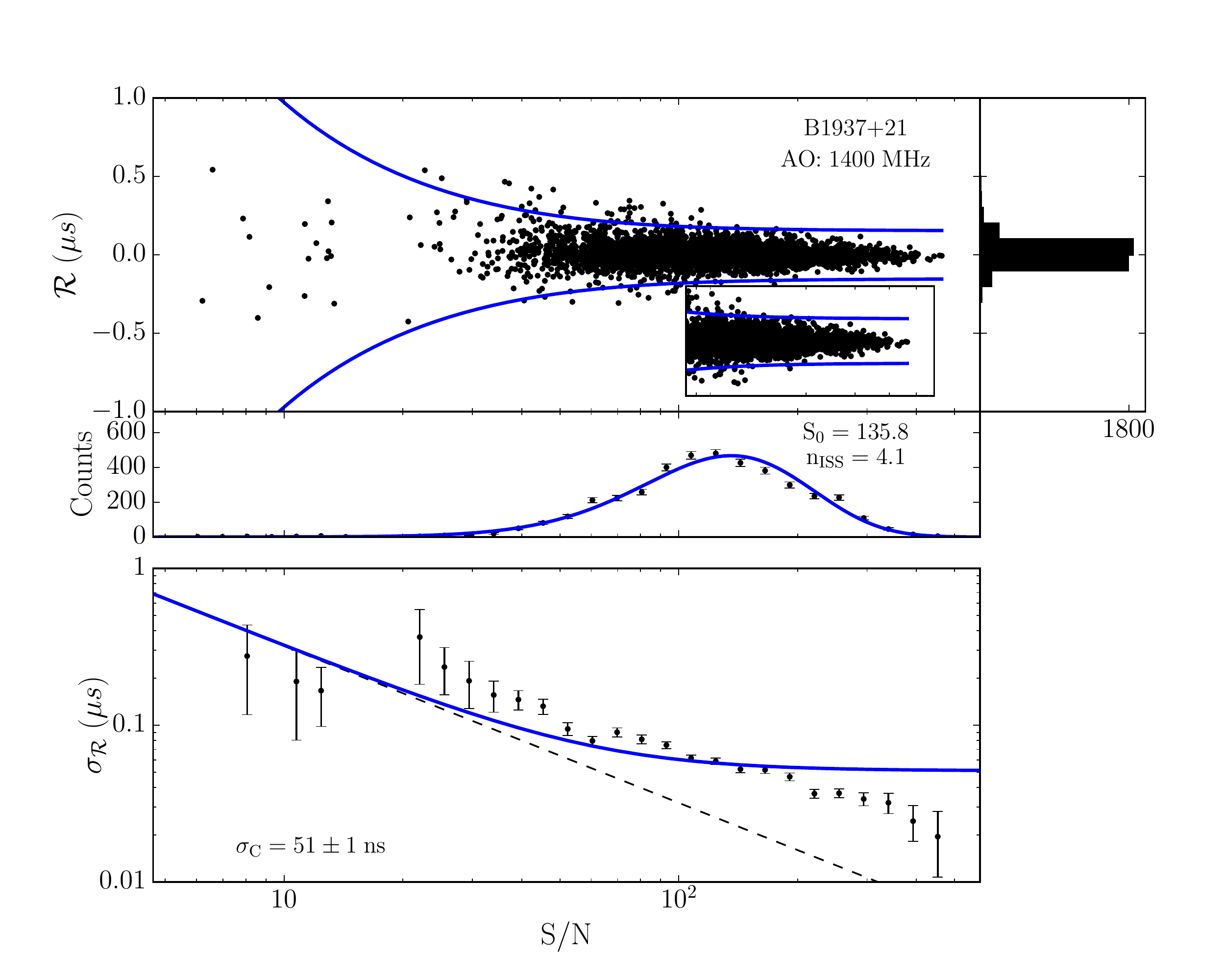}
  \caption{\footnotesize Analysis of residuals for PSR B1937+21 observed at 1400~MHz with AO. See the Figure \ref{fig:J1713+0747_L-wide_PUPPI} caption for more details.} 
\label{fig:B1937+21_L-wide_PUPPI}
\end{figure}

\begin{figure}[t!]
\hspace{-5ex}
\includegraphics[width=0.525\textwidth]{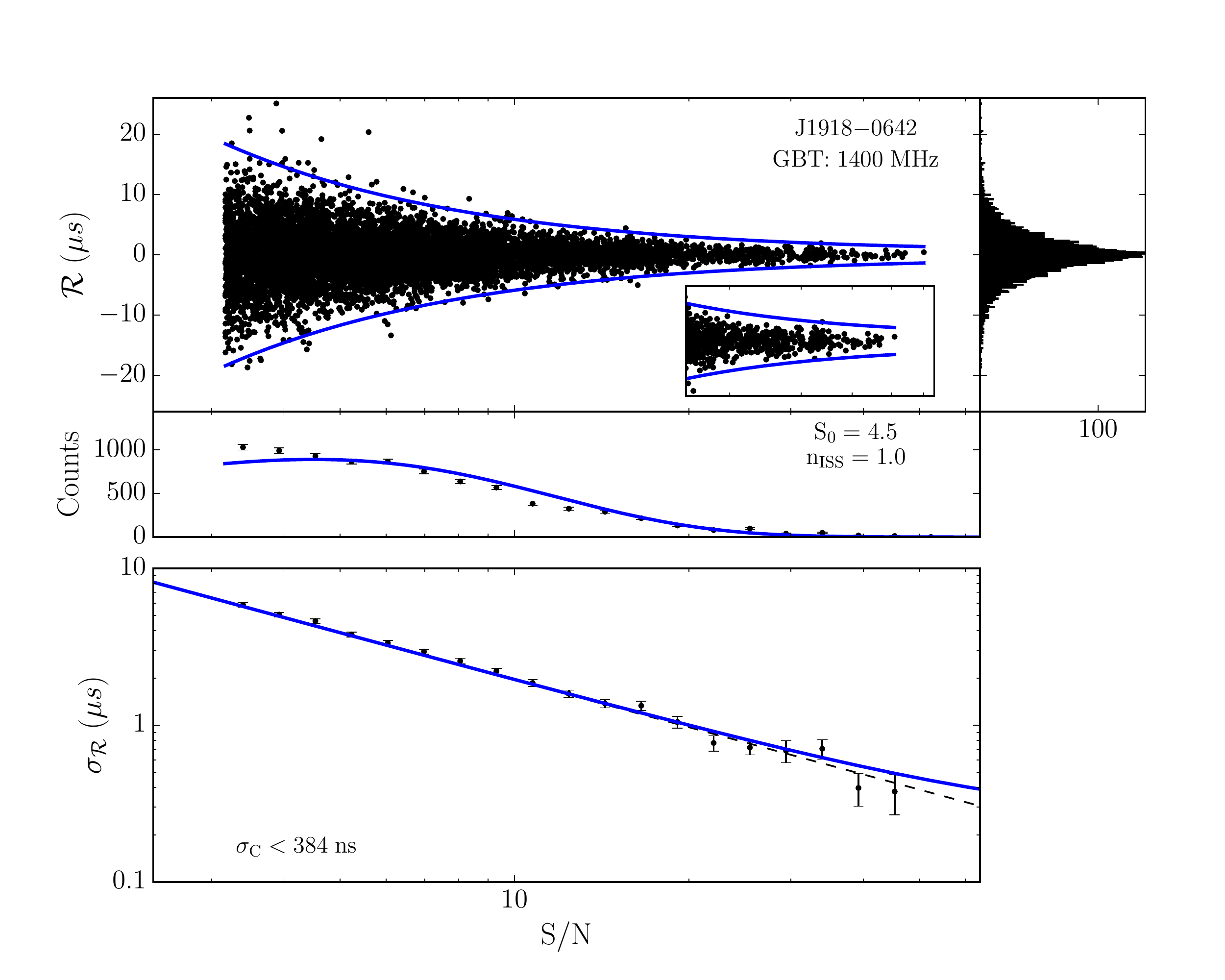}
  \caption{\footnotesize Analysis of residuals for PSR J1918$-$0642 observed at 1400~MHz with GBT. See the Figure \ref{fig:J1713+0747_L-wide_PUPPI} caption for more details. The 95\% upper limit on $\sigma_\C$ is shown in the bottom left of the bottom panel, with the corresponding $\sigma_\R$ in blue.}
\label{fig:J1918-0642_Rcvr1_2_GUPPI}
\end{figure}

\begin{figure*}[t!]
\epsscale{1.2} 
\begin{center}
\plotone{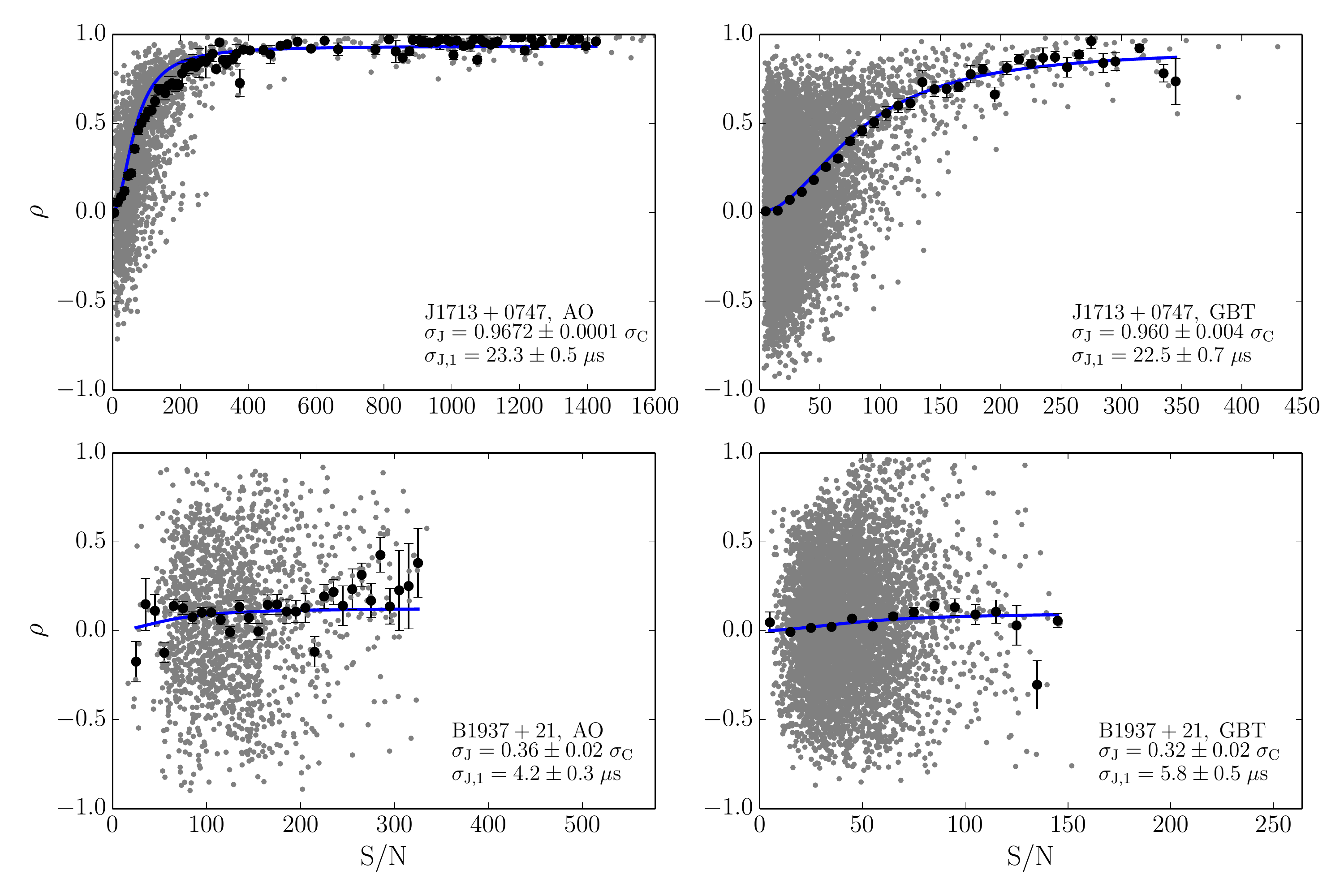}
  \caption{\footnotesize Correlation analysis for PSRs J1713+0747 (top) and B1937+21 (bottom) at AO (left) and GBT (right) for 1400~MHz band residuals. The gray points mark the correlation coefficient $\rho$ of two different sub-bands of residuals on a given epoch as a function of the geometric average of the mean $\SN$ of the pulse profiles for those sub-bands as $\left(\left<S_1\right>\left<S_2\right>\right)^{1/2}$. We show the median $\rho$ in bins of $\SN$ in black. The blue line marks the best-fit $\rho(S)$ to the black points.}
\label{fig:correlationsummary}
\end{center}
\end{figure*}

\begin{figure*}[t!]
\epsscale{1.2}
\begin{center}
\plotone{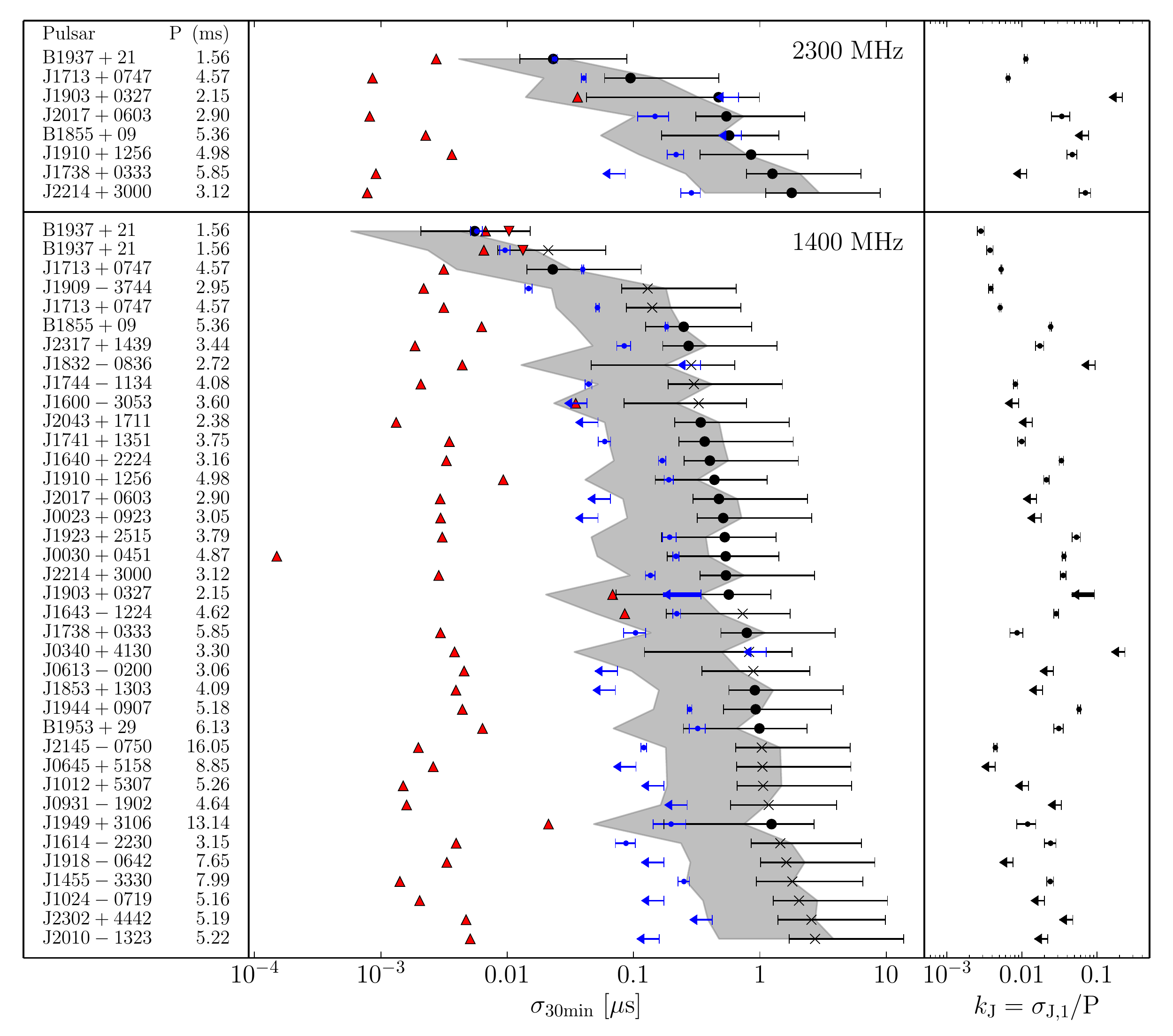}
  \caption{\footnotesize Summary of white-noise components for pulsars observed at the two highest frequency bands. The middle panel shows the three contributions, template-fitting noise as black circles (AO) or crosses (GBT), jitter noise as blue dots, and estimated DISS noise as red triangles. We observed PSRs J1713+0747 and B1937+21 with both telescopes and plot the separate analysis for each. The template-fitting and jitter contributions are for a 50 MHz bandwidth but scaled to a 30~min observating time. The gray bands represent the template-fitting noise scaled to the full receiver bandwidth to show the relative contribution with respect to jitter in a given NANOGrav observation. The DISS noise has been scaled to a 30~min observation and the appropriate total bandwidth for each band. Within each band, pulsars are ordered by increasing template-fitting noise (ordered by black points, not gray bands).  The rightmost panel shows the single-pulse RMS jitter divided by the period of the pulsar, $\kJ$. For PSR B1937+21, the upside-down triangles indicate the measured DISS noise from the correlation analysis (see \S\ref{subsec:corr_analysis}). The bold lines for PSR J1903+0327 at 1400~MHz indicates an upper limit on $\sigma_\C$ inconsistent with the estimate of $\sigma_{\DISS}$ (recall that $\sigma_{\DISS}$ has been scaled to the total bandwidth and a 30~min observing time and so appears smaller in the plot).}
\label{fig:summaryA}
\vspace{-2ex}
\end{center}
\end{figure*}

\begin{figure*}[t!]
\epsscale{1.2}
\begin{center}
\plotone{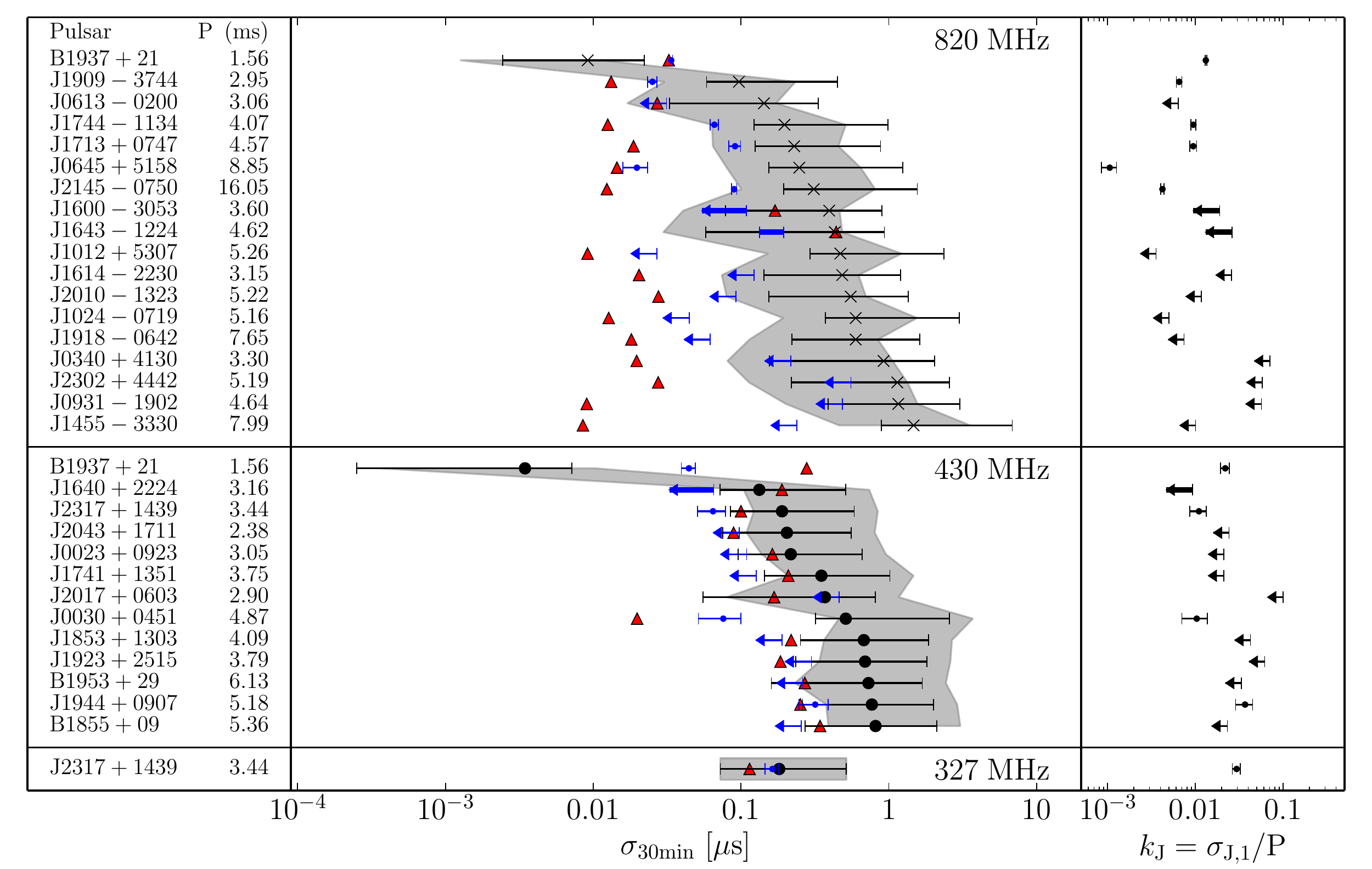}
  \caption{\footnotesize Summary of white-noise components for pulsars observed at the three lowest frequency bands. See the caption for Figure~\ref{fig:summaryA} for information. The bold upper limit lines for PSRs J1600$-$3053 and J1643$-$1224 at 820~MHz and J1640+2224 at 430~MHz, indicate an upper limit on $\sigma_\C$ inconsistent with the estimated $\sigma_{\DISS}$. }
\label{fig:summaryB}
\end{center}
\end{figure*}

The likelihood function can be calculated by combining Eqs.~\ref{eq:scintillation_pdf}, \ref{eq:fR}, and \ref{eq:scintillation_pdf_smin}, 
\ba
& & \Like(S_0,\niss,\sigma_\C | \{S_i,R_i\},\Smin) \nonumber \\
& & = \prod_i f_{\R,S}(\R_i,S_i | S_0,\niss,\sigma_\C,\Smin) \nonumber\\
& & = \prod_i f_{\R|S}(\R_i|S_i,\sigma_\C) f_S(S_i | S_0, \niss,\Smin),
\label{eq:likelihood}
\ea
where $i$ labels the individual residuals. We performed a grid search in the three-dimensional parameter space to estimate the values and uncertainties on the three model parameters. The likelihood function can be expressed as the product of individual likelihoods
\ba
& & \Like(S_0,\niss,\sigma_\C | \R,S,\Smin) \nonumber \\
& & = \Like(\sigma_\C | \R,S,\Smin) \Like(S_0,\niss|S,\Smin),
\label{eq:separable_likelihood}
\ea
so that we could perform the grid search in $\sigma_\C$ independently from the search in $S_0,\niss$ space. We limited our search in $\niss$ with a lower bound of 1 so that the minimum number of degrees of freedom across both the time and frequency dimensions is 2 \citep{cc1997}, or that each pulse must come from at least one ray path through the ISM. An F-test was used to determine the significance of $\sigma_\C$ with a significance value of 0.05 (i.e., $2\sigma$ significant). If not, we computed the 95\% upper limit on $\sigma_\C$.

Figure~\ref{fig:J1713+0747_L-wide_PUPPI} shows the results for one of NANOGrav's best-timed pulsars, PSR J1713+0747 observed at 1400~MHz at AO. The top panel shows the residuals $\R_i(S_i)$ with the $\pm 3\sigma_\R$ ranges plotted in the blue lines. At higher $\SN$, the RMS of the residuals asymptotes to a constant value, $\sigma_\C$, represented by the constant width scatter of points, and is indicative of jitter and scintillation noise and the $\SN$ regime over which they dominate the template-fitting error. A histogram of the residuals $R_i$ are shown at right with bins $\Delta \R = 0.1~\mu$s and a histogram  of $S_i$ with logarithmic bins of $\Delta \log_{10} S = 0.0625$ is shown below with Poisson uncertainties shown by the error bars. We plot $S f_S(S|S_0,\niss,\Smin)$ in the middle panel to properly compare the scaled PDF to the logarithmically binned histogram, with $S_0$ and $\niss$ determined in the ML analysis.

The bottom panel shows the RMS residual for the same logarithmic binning of the data. The dashed line shows the predicted RMS from template-fitting error only given by Eq.~\ref{eq:radiometer_noise}. We emphasize that the dashed line is not a fit to the data points in this plot. We see agreement between the dashed line and the points at low $\SN$ for most pulsars, which indicates that Eq.~\ref{eq:radiometer_noise} represents the template-fitting noise well. Deviation from the line can be explained by other systematic effects that can increase the variance, such as remaining RFI in the data. The blue line shows the best estimate $\sigma_\R(S)$ from the ML analysis. We note that the ML analysis is less susceptible to parameter mis-estimation from the effects of RFI in the data over a fit of Eq.~\ref{eq:one-parameter_model} to the RMS residual points because the ML analysis fits all of the data simultaneously.

Figures \ref{fig:J1909-3744_Rcvr1_2_GUPPI}-\ref{fig:J1918-0642_Rcvr1_2_GUPPI} show the same ML analysis for four other pulsars observed at 1400~MHz. While $\sigma_\R$ matches the data for PSR J1909$-$3744, the $\SN$ histogram does not match well with the data and the PDF of $\niss$ in the ML analysis peaks at the edge of the sampling space ($\niss = 1$), expected since $\Dnud = 39 \pm 14.7$~MHz and $\Dtd = 2258$~s for the pulsar at a reference frequency of 1500~MHz \citep{Keith+2013,Levin+2016}, of order the pulse channel bandwidth and typical total observation length. We see a similar result with PSR J2317+1439 though sparseness in the S/N histogram is a result of increased RFI excision for the pulsar. For PSR B1937+21, the $\SN$ histogram is well described by the result of the ML analysis. The remaining low $\SN$ residuals ($\SN \sim 10$) are spurious noise spikes that pass our $\SN$ cut criterion and lie close to the main pulse in phase. The narrowing of residuals at large S/N is not understood and may require further investigation of this pulsar. Lastly, we show the analysis for PSR J1918$-$0642 as a typical pulsar with an upper limit on $\sigma_\C$.

To measure jitter values, we estimated $\sigma_\DISS$ as described in the previous section for 50~MHz sub-bands and subintegrations of length $\tsub$ and then solved for $\sigma_\J$ given our measured $\sigma_\C$ (see Eq.~\ref{eq:one-parameter_model}). In several cases, the estimates of the scintillation noise from Eq.~\ref{eq:diss_noise} were larger than the $\sigma_\C$ estimated from the ML analysis, which is supposed to encapsulate all possible variance at high $\SN$. We employed a correlation analysis described in the next sub-section to separate the jitter and scintillation noise values for PSR B1937+21 at 1400 MHz, the only pulsar where the estimated $\sigma_{\DISS}$ is larger than $\sigma_\C$ and the S/N of the residuals is high enough to perform such an analysis. 

\subsection{Cross-Correlation Analysis Between Frequencies}
\label{subsec:corr_analysis}

Jitter causes simultaneously measured residuals at different frequencies to be correlated, which allows us to distinguish jitter noise from template-fitting noise. If the sub-band bandwidth is $\gtrsim \Dnud$, the residuals will not be correlated in frequency by DISS and we can distinguish jitter noise from scintillation noise as well. PSR B1937+21 has $\Dnud = 2.8 \pm 1.3$~MHz and $\Dtd = 327$~s at a reference frequency of 1500~MHz \citep{Keith+2013,Levin+2016} and therefore residuals with 50~MHz of bandwidth will be correlated in frequency due to jitter only. We find $\niss \approx 4$ for PSR B1937+21 observed at 1400~MHz at AO with 50~MHz sub-bands and $\sim 80$~s subintegrations. Therefore, the finite scintle effect is prominent and we expect scintillation noise to be large for this pulsar.



We let the total residual be the sum of the fluctuations from the three contributions to white noise,
\be
\R(\nu,t) = \R_{\SN}(\nu,t) + \R_\J(\nu,t) + \R_{\DISS}(\nu,t),
\label{eq:TOA_fluctuation}
\ee
where the subscripts denote the specific contribution. The cross-correlation coefficient between residuals from two sub-bands $\nu_i$ and $\nu_j$ is
\be
\left<\R(\nu_i,t)\R(\nu_j,t)\right> = \sigma_\J^2,
\label{eq:correlations}
\ee
where we assumed that the scintles are statistically independent between sub-bands for PSR B1937+21 and therefore do not correlate. The autocorrelation coefficient $\left<\R(\nu_i,t)^2\right>$ reduces to the variance in Eq.~\ref{eq:white_noise_model} plus cross terms that tend towards zero in the ensemble average limit. The correlation coefficient $\rho$ is the autocorrelation coefficient divided by the square root of the cross-correlation coefficients between sub-bands, which we assumed to be identical within a single band. Therefore,
\be
\rho(S) = \frac{\sigma_\J^2}{\sigma_{\R}^2(S)}
\label{eq:rho}
\ee
and is a function of pulse $\SN$. Since we calculated $\rho$ from residuals whose corresponding profiles differ in $\SN$, we took the average $\SN$ for the profiles in a given sub-band and used the geometric mean $\left(\left<S_i\right>\left<S_j\right>\right)^{1/2}$ as a proxy for pulse $\SN$.

Figure~\ref{fig:correlationsummary} shows the correlation coefficients (gray) as a function of $S$, computed over all epochs observed at 1400~MHz for PSRs J1713+0747 (top) and B1937+21 (bottom) at AO (left) and GBT (right). Since PSR J1713+0747 has $\sigma_\J \approx \sigma_\C$ and high-$\SN$, we show the results of our correlation analysis to demonstrate how the method performs before applying it to PSR B1937+21. The black points show the median $\rho$ with linear bins in $\SN$ each increasing by 10. The blue lines show the best fit of Eq.~\ref{eq:rho} to the black points via a grid search in $\sigma_\J$, holding $\sigma_\R$ fixed from the ML analysis. Each panel shows $\sigma_\J$ as a fraction of $\sigma_\C$ as well as the single-pulse RMS $\sigma_{\J,1}$, which accounts for the differences in $\tsub$ between telescopes. The errors include both errors on $\sigma_\C$ and errors from the fit.

For PSR J1713+0747, we find consistency of $\sigma_{\J,1}$ between AO and GBT, with $\sigma_{\J,1} = 23.3 \pm 0.5~\mu$s, $22.5 \pm 0.7~\mu$s, respectively, which demonstrate a good check of the methods used in this paper. These values are generally consistent with, though somewhat lower than, measurements reported elsewhere. \citet{Dolch+2014} report a measurement of jitter though their method includes the contribution from $\sigma_{\DISS}$. They find $\sigma_{\C,1} = 27.0 \pm 3.3~\mu$s, though we note that the $\sigma_{\DISS}$ contribution is not well-defined at the single-pulse level. \citet{sc2012} find $\sigma_{\J,1} \approx 26~\mu$s from a cross-correlation analysis between frequency bands. \citet{sod+2014} find $\sigma_{\C,1} = 31.1 \pm 0.7~\mu$s (again including the contribution of $\sigma_{\DISS}$) by adding Gaussian noise to the template, generating residuals, and subtracting the quadrature difference from the observed residuals. Even accounting for the small contribution from $\sigma_{\DISS}$, their measurement formally disagrees with ours for reasons that are uncertain.

For PSR B1937+21, $\sigma_{\DISS}$ is comparable to the predicted values from the scaling relation (Eq.~\ref{eq:diss_noise}) for both AO and GBT at 1400~MHz. Differences between the estimated $\sigma_{\J,1}$ between telescopes come from differences in the estimated $\sigma_\C$ whereas the ratio of RMSs $\sigma_\J/\sigma_\C$ is consistent between the measurements at both observatories. Since the GUPPI observations span more years than the PUPPI observations ($\sim 3.6$~yr versus $\sim 1.6$~yr, respectively), if the scintillation parameters differed in the first half of the GUPPI observations than the second half when PUPPI ran in coincidence, then the average $\sigma_{\DISS}$ would differ between the two sets of observations. The small scintle size at 1400~MHz means that we are unable to study the scintillation properties of this pulsar with the current NANOGrav data set.

\section{Summary Results}
\label{sec:analysis_summary}

Figures \ref{fig:summaryA} and \ref{fig:summaryB} show the results for the three white-noise contributions to the timing residuals per frequency band per pulsar. We performed the ML analysis independently for observations of PSRs J1713+0747 and B1937+21, which were observed at both telescopes. For each frequency band, the pulsars are ordered in increasing amounts of template-fitting noise. Template-fitting noise values are calculated using Eq.~\ref{eq:radiometer_noise} and using the median and 68.3\% confidence limits from the PDFs of $\SN$ for each pulsar. Jitter values are also 68.3\% confidence intervals or upper limits at the 95\% level. DISS noise is calculated through scattering measurements as discussed in \S\ref{sec:observations} and according to Eq.~\ref{eq:diss_noise}. We scale the observation time $T$ to 30~min and the bandwidth $B$ equal to that of each receiver in NG9 \citep[see Table 1 of][]{NG9yr}.

To compare numbers expected over the length of a typical NANOGrav observation, we scaled all three contributions to 30~min. We multiplied the mean $\SN$, $S_0$ (see Eq.~\ref{eq:scintillation_pdf}), by a factor of $\sqrt{30~\mathrm{min}/\tsub}$, where $\tsub$ is the subintegration time for either GBT ($\sim 120$~s) or AO, ($\sim 80$~s) to find the 30~min $\SN$ for use in Eq.~\ref{eq:radiometer_noise}. Because the scintillation timescales are of the order of the typical observation length or longer for most of these MSPs \citep{Levin+2016}, the simple scaling relation of Eq.~\ref{eq:N_DISS} will hold on average though not exactly since the number of scintles in the time dimension is restricted. The scintillation noise term was scaled up in time and frequency using Eq.~\ref{eq:diss_noise}. The gray band shows the template-fitting error scaled to the full bandwidth $B$ by a factor of $\sqrt{50~\mathrm{MHz}/B}$. The rightmost panel shows the jitter parameter $\kJ = \sigma_{\J,1}/P$.

The raw values from our analysis are reported in Table~\ref{table:raw}. In Table~\ref{table:summary}, we convert all three white-noise contribution measurements to 30~min TOA uncertainties and rank the pulsars according to each contribution and to the total white noise per frequency band (thus matching Figures~\ref{fig:summaryA} and \ref{fig:summaryB}).

\subsection{Pulse Jitter Statistics}
\label{sec:jitter_summary}

The preceding analysis provides detections of $\sigma_\J$ for over half of the NANOGrav pulsars for the 1400~MHz band. This large sample allows us to examine the statistics of the jitter distribution. We use the jitter parameter $\kJ$ to compare pulsars, since it is independent of the pulse period.

\renewcommand{\thefootnote}{\alph{footnote}}
\LongTables

\begin{deluxetable*}{ccccccccccccc}
\tablecolumns{13}
\tablecaption{Maximum Likelihood Analysis Output}
\tablehead{
\colhead{Pulsar} & \colhead{Period} & \colhead{DM} & \colhead{Telescope} & \colhead{Frequency} & \colhead{$W_{\mathrm{eff}}$} & \colhead{$W_{50}$} & \colhead{$t_{\mathrm{sub}^{\rm a}}$} & \colhead{$S_0$} & \colhead{$n_{\mathrm{ISS}}$} & \colhead{$\sigma_{\mathrm{C}}$} & \colhead{$+\sigma_{\sigma_{\mathrm{C}}}$} & \colhead{$-\sigma_{\sigma_{\mathrm{C}}}$} \\
\colhead{} & \colhead{(ms)} & \colhead{(pc cm$^{-3}$)} & \colhead{} & \colhead{(MHz)} & \colhead{($\mathrm{\mu s}$)} & \colhead{($\mathrm{\mu s}$)} & \colhead{($\mathrm{s}$)} & \colhead{} & \colhead{} & \colhead{($\mathrm{ns}$)} & \colhead{($\mathrm{ns}$)} & \colhead{($\mathrm{ns}$)} 
}
\startdata
J0023+0923 & 3.05 & 14.33 & AO & 430 & 335 & 72 & 84.6 & 8.4 & 2.7 & $<$363 & - & -\\*
 & & & AO & 1400 & 430 & 201 & 84.6 & 5.8 & 1.0 & $<$273 & - & -\\*
J0030+0451 & 4.87 & 4.33 & AO & 430 & 705 & 643 & 85.9 & 9.6 & 1.0 & 380 & 120 & 122\\*
 & & & AO & 1400 & 540 & 505 & 85.9 & 5.2 & 5.1 & 1328 & 71 & 71\\*
J0340+4130 & 3.30 & 49.58 & GBT & 800 & 545 & 189 & 126.0 & 3.5 & 27.6 & $<$999 & - & -\\*
 & & & GBT & 1400 & 515 & 213 & 126.0 & 3.7 & 37.0 & $<$3301 & - & -\\*
J0613$-$0200 & 3.06 & 38.78 & GBT & 800 & 250 & 66 & 126.5 & 10.5 & 13.7 & $<$97 & - & -\\*
 & & & GBT & 1400 & 331 & 274 & 126.5 & 2.4 & 3.7 & $<$328 & - & -\\*
J0645+5158 & 8.85 & 18.25 & GBT & 800 & 590 & 70 & 128.8 & 20.3 & 1.0 & 85 & 16 & 16\\*
 & & & GBT & 1400 & 627 & 117 & 128.8 & 5.1 & 1.0 & $<$268 & - & -\\*
J0931$-$1902 & 4.64 & 41.49 & GBT & 800 & 669 & 296 & 126.5 & 3.6 & 5.5 & $<$1323 & - & -\\*
 & & & GBT & 1400 & 466 & 348 & 126.5 & 2.8 & 1.9 & $<$786 & - & -\\*
J1012+5307 & 5.26 & 9.02 & GBT & 800 & 667 & 691 & 128.8 & 12.1 & 1.0 & $<$102 & - & -\\*
 & & & GBT & 1400 & 634 & 584 & 128.8 & 5.1 & 1.0 & $<$343 & - & -\\*
J1024$-$0719 & 5.16 & 6.49 & GBT & 800 & 553 & 338 & 128.8 & 7.9 & 1.0 & $<$138 & - & -\\*
 & & & GBT & 1400 & 574 & 141 & 128.8 & 2.4 & 1.0 & $<$541 & - & -\\*
J1455$-$3330 & 7.99 & 13.57 & GBT & 800 & 956 & 440 & 121.7 & 5.2 & 1.1 & $<$541 & - & -\\*
 & & & GBT & 1400 & 996 & 207 & 121.7 & 3.9 & 1.7 & 1542 & 162 & 161\\*
J1600$-$3053 & 3.60 & 52.33 & GBT & 800 & 485 & 102 & 122.5 & 7.2 & 19.2 & $<$277 & - & -\\*
 & & & GBT & 1400 & 424 & 70 & 122.5 & 7.7 & 10.6 & $<$236 & - & -\\*
J1614$-$2230 & 3.15 & 34.50 & GBT & 800 & 449 & 109 & 122.5 & 5.6 & 7.5 & $<$348 & - & -\\*
 & & & GBT & 1400 & 391 & 84 & 122.5 & 2.1 & 1.2 & 385 & 71 & 68\\*
J1640+2224 & 3.16 & 18.43 & AO & 430 & 383 & 96 & 84.6 & 17.5 & 1.5 & $<$135 & - & -\\*
 & & & AO & 1400 & 465 & 220 & 84.6 & 8.0 & 1.0 & 648 & 43 & 43\\*
J1643$-$1224 & 4.62 & 62.41 & GBT & 800 & 1040 & 390 & 122.5 & 14.0 & 46.7 & 555 & 106 & 101\\*
 & & & GBT & 1400 & 973 & 315 & 122.5 & 7.9 & 11.5 & 899 & 65 & 64\\*
J1713+0747 & 4.57 & 15.99 & AO & 1400 & 539 & 110 & 82.0 & 159.9 & 1.0 & 180 & 4 & 3\\*
 & & & AO & 2300 & 512 & 104 & 82.0 & 36.8 & 1.0 & 223 & 10 & 10\\*
 & & & GBT & 800 & 694 & 170 & 121.7 & 22.1 & 1.5 & 268 & 24 & 24\\*
 & & & GBT & 1400 & 533 & 109 & 121.7 & 31.4 & 1.0 & 143 & 4 & 4\\*
J1738+0333 & 5.85 & 33.77 & AO & 1400 & 643 & 120 & 83.4 & 5.6 & 1.0 & 421 & 84 & 83\\*
 & & & AO & 2300 & 696 & 118 & 83.4 & 3.8 & 1.0 & $<$472 & - & -\\*
J1741+1351 & 3.75 & 24.19 & AO & 430 & 458 & 109 & 85.0 & 7.0 & 3.2 & $<$475 & - & -\\*
 & & & AO & 1400 & 390 & 86 & 85.0 & 7.4 & 1.0 & 247 & 28 & 28\\*
J1744$-$1134 & 4.07 & 3.14 & GBT & 800 & 513 & 147 & 121.7 & 21.6 & 1.0 & 225 & 15 & 15\\*
 & & & GBT & 1400 & 511 & 137 & 121.7 & 14.1 & 1.0 & 193 & 12 & 12\\*
 & & & GBT & 1400 & 187 & 49 & 121.7 & 3.8 & 30.6 & $<$1006 & - & -\\*
J1853+1303 & 4.09 & 30.57 & AO & 430 & 486 & 606 & 83.4 & 3.7 & 4.2 & $<$1035 & - & -\\*
 & & & AO & 1400 & 346 & 125 & 83.4 & 2.6 & 1.0 & $<$451 & - & -\\*
B1855+09 & 5.36 & 13.30 & AO & 430 & 796 & 653 & 85.2 & 5.0 & 5.6 & $<$888 & - & -\\*
 & & & AO & 1400 & 750 & 518 & 85.2 & 17.4 & 1.9 & 1025 & 25 & 25\\*
 & & & AO & 2300 & 716 & 485 & 85.2 & 6.3 & 7.7 & $<$2722 & - & -\\*
J1903+0327 & 2.15 & 297.54 & AO & 1400 & 405 & 195 & 82.6 & 3.4 & 51.3 & $<$760 & - & -\\*
 & & & AO & 2300 & 327 & 99 & 82.6 & 3.3 & 108.8 & $<$2023 & - & -\\*
J1909$-$3744 & 2.95 & 10.39 & GBT & 800 & 279 & 53 & 121.7 & 23.1 & 1.1 & 99 & 7 & 7\\*
 & & & GBT & 1400 & 261 & 41 & 121.7 & 16.7 & 1.0 & 56 & 4 & 4\\*
J1910+1256 & 4.98 & 38.06 & AO & 1400 & 634 & 133 & 82.0 & 7.3 & 5.3 & 823 & 67 & 67\\*
 & & & AO & 2300 & 574 & 108 & 82.0 & 3.5 & 3.6 & 1819 & 273 & 273\\*
J1918$-$0642 & 7.65 & 26.59 & GBT & 800 & 979 & 184 & 121.7 & 10.2 & 4.3 & $<$377 & - & -\\*
 & & & GBT & 1400 & 879 & 151 & 121.7 & 4.5 & 1.0 & $<$384 & - & -\\*
J1923+2515 & 3.79 & 18.86 & AO & 430 & 448 & 175 & 83.9 & 3.3 & 5.3 & $<$1319 & - & -\\*
 & & & AO & 1400 & 534 & 146 & 83.9 & 5.1 & 6.4 & 1355 & 168 & 168\\*
B1937+21 & 1.56 & 71.02 & AO & 430 & 190 & 63 & 84.1 & 263.1 & 170.5 & 448 & 48 & 50\\*
 & & & AO & 1400 & 145 & 37 & 84.1 & 135.8 & 4.1 & 51 & 1 & 1\\*
 & & & AO & 2300 & 147 & 36 & 84.1 & 38.4 & 1.5 & 76 & 4 & 4\\*
 & & & GBT & 800 & 153 & 54 & 120.9 & 98.6 & 9.7 & 128 & 3 & 3\\*
 & & & GBT & 1400 & 146 & 37 & 120.9 & 43.6 & 3.5 & 66 & 2 & 1\\*
J1944+0907 & 5.18 & 24.34 & AO & 430 & 1120 & 500 & 83.4 & 7.4 & 5.3 & 1524 & 341 & 336\\*
 & & & AO & 1400 & 949 & 364 & 83.4 & 6.3 & 1.4 & 2354 & 99 & 99\\*
J1949+3106 & 13.14 & 164.13 & AO & 1400 & 916 & 142 & 80.8 & 3.5 & 41.8 & 1996 & 608 & 561\\*
B1953+29 & 6.13 & 104.58 & AO & 430 & 1293 & 481 & 84.6 & 8.7 & 15.1 & $<$1505 & - & -\\*
 & & & AO & 1400 & 823 & 224 & 84.6 & 4.1 & 11.2 & 1615 & 235 & 230\\*
J2010$-$1323 & 5.22 & 22.16 & GBT & 800 & 499 & 240 & 121.7 & 5.4 & 8.6 & $<$342 & - & -\\*
 & & & GBT & 1400 & 527 & 247 & 121.7 & 1.6 & 1.0 & $<$631 & - & -\\*
J2017+0603 & 2.90 & 23.92 & AO & 430 & 323 & 62 & 83.4 & 4.2 & 36.1 & $<$1430 & - & -\\*
 & & & AO & 1400 & 242 & 64 & 83.4 & 3.5 & 1.0 & $<$223 & - & -\\*
 & & & AO & 2300 & 234 & 61 & 83.4 & 2.7 & 1.3 & 578 & 162 & 157\\*
J2043+1711 & 2.38 & 20.71 & AO & 430 & 222 & 35 & 83.5 & 5.6 & 4.3 & $<$269 & - & -\\*
 & & & AO & 1400 & 178 & 21 & 83.5 & 3.6 & 1.0 & $<$145 & - & -\\*
J2145$-$0750 & 16.05 & 9.01 & GBT & 800 & 1826 & 395 & 121.7 & 48.7 & 1.0 & 778 & 32 & 32\\*
 & & & GBT & 1400 & 1823 & 339 & 121.7 & 14.6 & 1.0 & 815 & 40 & 40\\
J2214+3000 & 3.12 & 22.56 & AO & 1400 & 562 & 181 & 82.0 & 7.1 & 1.0 & 682 & 60 & 60\\*
 & & & AO & 2300 & 551 & 180 & 82.0 & 2.1 & 1.0 & 1345 & 236 & 237\\*
J2302+4442 & 5.19 & 13.76 & GBT & 800 & 608 & 345 & 128.8 & 3.2 & 20.9 & $<$1588 & - & -\\*
 & & & GBT & 1400 & 682 & 347 & 128.8 & 2.0 & 1.5 & $<$1307 & - & -\\*
J2317+1439 & 3.44 & 21.90 & AO & 327 & 395 & 152 & 85.7 & 11.6 & 3.5 & 677 & 71 & 72\\*
 & & & AO & 430 & 392 & 169 & 85.7 & 11.4 & 2.6 & 266 & 58 & 57\\*
 & & & AO & 1400 & 376 & 152 & 85.7 & 9.6 & 1.0 & 378 & 47 & 48
\enddata
\label{table:raw}
\tablenotetext{a}{Median subintegration length}
\end{deluxetable*}

\begin{deluxetable*}{cccccccccccc}
\tablecolumns{12}
\tablecaption{Summary of White Noise Contributions}
\tablehead{
\colhead{Pulsar} & \colhead{Period} & \colhead{DM} & \colhead{Telescope} & \colhead{$\sigma_{\mathrm{S/N}}$} & \colhead{$\sigma_{\mathrm{J}}$} & \colhead{$\sigma_{\mathrm{DISS}}$} & \colhead{$\sigma_{\mathcal{R}}$} & \colhead{Rank S/N} & \colhead{Rank J} & \colhead{Rank DISS} & \colhead{Rank Total}\\
\colhead{} & \colhead{(ms)} & \colhead{(pc cm$^{-3}$)}  & \colhead{}  & \colhead{(ns)} & \colhead{(ns)} & \colhead{(ns)} & \colhead{(ns)} & \colhead{}  & \colhead{}  & \colhead{}  & \colhead{} 
}
\startdata
\cutinhead{327 MHz}\\* 
J2317+1439 & 3.44 & 21.90 & AO & 181 & 163 & 114 & 269 & 1 & 1 & 1 & 1\\*
\cutinhead{430 MHz}\\* 
J0023+0923 & 3.05 & 14.33 & AO & 313 & $<$90 & 163 & $<$364 & 5 & 6 & 4 & 5\\*
J0030+0451 & 4.87 & 4.33 & AO & 737 & 75 & 19 & 742 & 8 & 4 & 1 & 8\\*
J1640+2224 & 3.16 & 18.43 & AO & 191 & 49$^{\rm a}$ & 189 & 198$^{\rm a}$ & 2 & 2 & 7 & 1\\*
J1741+1351 & 3.75 & 24.19 & AO & 504 & $<$105 & 208 & $<$556 & 6 & 7 & 8 & 6\\*
J1853+1303 & 4.09 & 30.57 & AO & 977 & $<$158 & 218 & $<$1014 & 9 & 8 & 9 & 9\\*
B1855+09 & 5.36 & 13.30 & AO & 1173 & $<$213 & 342 & $<$1240 & 13 & 9 & 13 & 13\\*
J1923+2515 & 3.79 & 18.86 & AO & 996 & $<$249 & 184 & $<$1043 & 10 & 11 & 6 & 10\\*
B1937+21 & 1.56 & 71.02 & AO & 5.0 & 44 & 278 & 281 & 1 & 1 & 12 & 2\\*
J1944+0907 & 5.18 & 24.34 & AO & 1108 & 317 & 252 & 1179 & 12 & 12 & 10 & 12\\*
B1953+29 & 6.13 & 104.58 & AO & 1051 & $<$217 & 270 & $<$1106 & 11 & 10 & 11 & 11\\*
J2017+0603 & 2.90 & 23.92 & AO & 533 & $<$384 & 167 & $<$678 & 7 & 13 & 5 & 7\\*
J2043+1711 & 2.38 & 20.71 & AO & 295 & $<$81 & 89 & $<$318 & 4 & 5 & 2 & 4\\*
J2317+1439 & 3.44 & 21.90 & AO & 273 & 64 & 99 & 298 & 3 & 3 & 3 & 3\\*
\cutinhead{820~MHz}\\* 
J0340+4130 & 3.30 & 49.58 & GBT & 478 & $<$181 & 19 & $<$511 & 15 & 15 & 11 & 15\\*
J0613$-$0200 & 3.06 & 38.78 & GBT & 74 & $<$26 & 27 & $<$83 & 3 & 4 & 13 & 3\\*
J0645+5158 & 8.85 & 18.25 & GBT & 128 & 19 & 14 & 130 & 6 & 1 & 8 & 5\\*
J0931$-$1902 & 4.64 & 41.49 & GBT & 600 & $<$403 & 9.0 & $<$723 & 17 & 17 & 2 & 16\\*
J1012+5307 & 5.26 & 9.02 & GBT & 243 & $<$22 & 9.2 & $<$244 & 10 & 2 & 3 & 9\\*
J1024$-$0719 & 5.16 & 6.49 & GBT & 309 & $<$37 & 12 & $<$311 & 13 & 6 & 6 & 13\\*
J1455$-$3330 & 7.99 & 13.57 & GBT & 762 & $<$199 & 8.5 & $<$788 & 18 & 16 & 1 & 18\\*
J1600$-$3053 & 3.60 & 52.33 & GBT & 204 & $<$81$^{\rm a}$ & 169 & $<$220$^{\rm a}$ & 8 & 10 & 17 & 8\\*
J1614$-$2230 & 3.15 & 34.50 & GBT & 250 & $<$101 & 20 & $<$271 & 11 & 13 & 12 & 10\\*
J1643$-$1224 & 4.62 & 62.41 & GBT & 223 & 162$^{\rm a}$ & 440 & 276$^{\rm a}$ & 9 & 14 & 18 & 11\\*
J1713+0747 & 4.57 & 15.99 & GBT & 118 & 91 & 18 & 150 & 5 & 12 & 10 & 6\\*
J1744$-$1134 & 4.07 & 3.14 & GBT & 102 & 66 & 12 & 122 & 4 & 8 & 5 & 4\\*
J1909$-$3744 & 2.95 & 10.39 & GBT & 50 & 25 & 13 & 57 & 2 & 3 & 7 & 2\\*
J1918$-$0642 & 7.65 & 26.59 & GBT & 309 & $<$51 & 18 & $<$314 & 14 & 7 & 9 & 14\\*
B1937+21 & 1.56 & 71.02 & GBT & 4.8 & 33 & 32 & 47 & 1 & 5 & 16 & 1\\*
J2010$-$1323 & 5.22 & 22.16 & GBT & 286 & $<$77 & 27 & $<$297 & 12 & 9 & 15 & 12\\*
J2145$-$0750 & 16.05 & 9.01 & GBT & 161 & 89 & 12 & 184 & 7 & 11 & 4 & 7\\*
J2302+4442 & 5.19 & 13.76 & GBT & 591 & $<$461 & 27 & $<$750 & 16 & 18 & 14 & 17\\*
\cutinhead{1400 MHz}\\* 
J0023+0923 & 3.05 & 14.33 & AO & 142 & $<$43 & 3.0 & $<$149 & 16 & 6 & 15 & 12\\*
J0030+0451 & 4.87 & 4.33 & AO & 149 & 216 & 0.2 & 263 & 18 & 29 & 1 & 21\\*
J0340+4130 & 3.30 & 49.58 & GBT & 229 & $<$935 & 3.8 & $<$963 & 23 & 38 & 22 & 38\\*
J0613$-$0200 & 3.06 & 38.78 & GBT & 247 & $<$62 & 4.6 & $<$255 & 24 & 13 & 27 & 19\\*
J0645+5158 & 8.85 & 18.25 & GBT & 292 & $<$87 & 2.6 & $<$305 & 29 & 16 & 11 & 24\\*
J0931$-$1902 & 4.64 & 41.49 & GBT & 327 & $<$221 & 1.6 & $<$394 & 31 & 31 & 5 & 29\\*
J1012+5307 & 5.26 & 9.02 & GBT & 295 & $<$145 & 1.5 & $<$329 & 30 & 22 & 4 & 27\\*
J1024$-$0719 & 5.16 & 6.49 & GBT & 568 & $<$145 & 2.0 & $<$587 & 36 & 23 & 8 & 35\\*
J1455$-$3330 & 7.99 & 13.57 & GBT & 503 & 250 & 1.4 & 562 & 35 & 32 & 3 & 34\\
J1600$-$3053 & 3.60 & 52.33 & GBT & 91 & $<$35 & 34 & $<$104 & 10 & 4 & 36 & 7\\*
J1614$-$2230 & 3.15 & 34.50 & GBT & 404 & 87 & 3.9 & 413 & 33 & 15 & 24 & 31\\*
J1640+2224 & 3.16 & 18.43 & AO & 112 & 168 & 3.3 & 202 & 13 & 24 & 19 & 14\\*
J1643$-$1224 & 4.62 & 62.41 & GBT & 204 & 219 & 85 & 311 & 21 & 30 & 38 & 25\\*
J1713+0747 & 4.57 & 15.99 & AO & 6.4 & 39 & 3.2 & 40 & 3 & 5 & 17 & 4\\*
 & & & GBT & 39 & 51 & 3.2 & 65 & 5 & 9 & 18 & 5\\*
J1738+0333 & 5.85 & 33.77 & AO & 219 & 103 & 3.0 & 243 & 22 & 17 & 14 & 18\\*
J1741+1351 & 3.75 & 24.19 & AO & 102 & 59 & 3.5 & 117 & 12 & 11 & 21 & 10\\*
J1744$-$1134 & 4.08 & 3.14 & GBT & 83 & 44 & 2.1 & 94 & 9 & 8 & 9 & 6\\*
J1832$-$0836 & 2.72 & 28.18 & GBT & 79 & $<$281 & 4.4 & $<$292 & 8 & 35 & 25 & 22\\*
J1853+1303 & 4.09 & 30.57 & AO & 254 & $<$59 & 3.9 & $<$261 & 25 & 12 & 23 & 20\\*
B1855+09 & 5.36 & 13.30 & AO & 69 & 182 & 6.3 & 195 & 6 & 25 & 30 & 13\\*
J1903+0327 & 2.15 & 297.54 & AO & 158 & 257$^{\rm a}$ & 68 & 301$^{\rm a}$ & 20 & 33 & 37 & 23\\*
J1909$-$3744 & 2.95 & 10.39 & GBT & 36 & 14 & 2.2 & 39 & 4 & 3 & 10 & 3\\*
J1910+1256 & 4.98 & 38.06 & AO & 121 & 190 & 9.3 & 226 & 14 & 26 & 32 & 16\\*
J1918$-$0642 & 7.65 & 26.59 & GBT & 451 & $<$144 & 3.3 & $<$474 & 34 & 21 & 20 & 33\\*
J1923+2515 & 3.79 & 18.86 & AO & 147 & 193 & 3.1 & 242 & 17 & 27 & 16 & 17\\*
B1937+21 & 1.56 & 71.02 & AO & 1.6 & 5.7 & 10 & 11 & 1 & 1 & 33 & 1\\*
 & & & GBT & 5.9 & 9.6 & 13 & 17 & 2 & 2 & 34 & 2\\*
J1944+0907 & 5.18 & 24.34 & AO & 258 & 277 & 4.4 & 379 & 26 & 34 & 26 & 28\\*
J1949+3106 & 13.14 & 164.13 & AO & 344 & 198 & 21 & 398 & 32 & 28 & 35 & 30\\*
B1953+29 & 6.13 & 104.58 & AO & 276 & 321 & 6.4 & 424 & 27 & 36 & 31 & 32\\*
J2010$-$1323 & 5.22 & 22.16 & GBT & 762 & $<$133 & 5.1 & $<$773 & 38 & 19 & 29 & 36\\*
J2017+0603 & 2.90 & 23.92 & AO & 132 & $<$54 & 3.0 & $<$143 & 15 & 10 & 13 & 11\\*
J2043+1711 & 2.38 & 20.71 & AO & 94 & $<$43 & 1.3 & $<$104 & 11 & 7 & 2 & 8\\*
J2145$-$0750 & 16.05 & 9.01 & GBT & 288 & 120 & 2.0 & 312 & 28 & 18 & 7 & 26\\*
J2214+3000 & 3.12 & 22.56 & AO & 150 & 136 & 2.9 & 202 & 19 & 20 & 12 & 15\\*
J2302+4442 & 5.19 & 13.76 & GBT & 713 & $<$349 & 4.7 & $<$794 & 37 & 37 & 28 & 37\\*
J2317+1439 & 3.44 & 21.90 & AO & 76 & 84 & 1.9 & 113 & 7 & 14 & 6 & 9\\*
\cutinhead{2300 MHz}\\* 
J1713+0747 & 4.57 & 15.99 & AO & 31 & 40 & 0.9 & 51 & 2 & 2 & 3 & 2\\*
J1738+0333 & 5.85 & 33.77 & AO & 414 & $<$71 & 0.9 & $<$420 & 7 & 3 & 4 & 5\\*
B1855+09 & 5.36 & 13.30 & AO & 188 & $<$592 & 2.3 & $<$621 & 5 & 8 & 5 & 7\\*
J1903+0327 & 2.15 & 297.54 & AO & 155 & $<$563 & 36 & $<$585 & 3 & 7 & 8 & 6\\*
J1910+1256 & 4.98 & 38.06 & AO & 280 & 217 & 3.6 & 354 & 6 & 5 & 7 & 4\\*
B1937+21 & 1.56 & 71.02 & AO & 7.6 & 23 & 2.7 & 25 & 1 & 1 & 6 & 1\\*
J2017+0603 & 2.90 & 23.92 & AO & 179 & 148 & 0.8 & 232 & 4 & 4 & 2 & 3\\*
J2214+3000 & 3.12 & 22.56 & AO & 588 & 287 & 0.8 & 654 & 8 & 6 & 1 & 8
\enddata
\label{table:summary}
\tablenotetext{a}{When the estimated $\sigma_{\DISS}$ is larger than the measured $\sigma_\C$, entries for $\sigma_\J$ are replaced by the values for $\sigma_\C$. The total residual RMS $\sigma_\R$ is set equal to $\sqrt{\sigma_\C^2 + \sigma_{\SN}^2}$}
\end{deluxetable*}

\begin{figure}[t!]
\begin{center}
\includegraphics[width=0.5\textwidth]{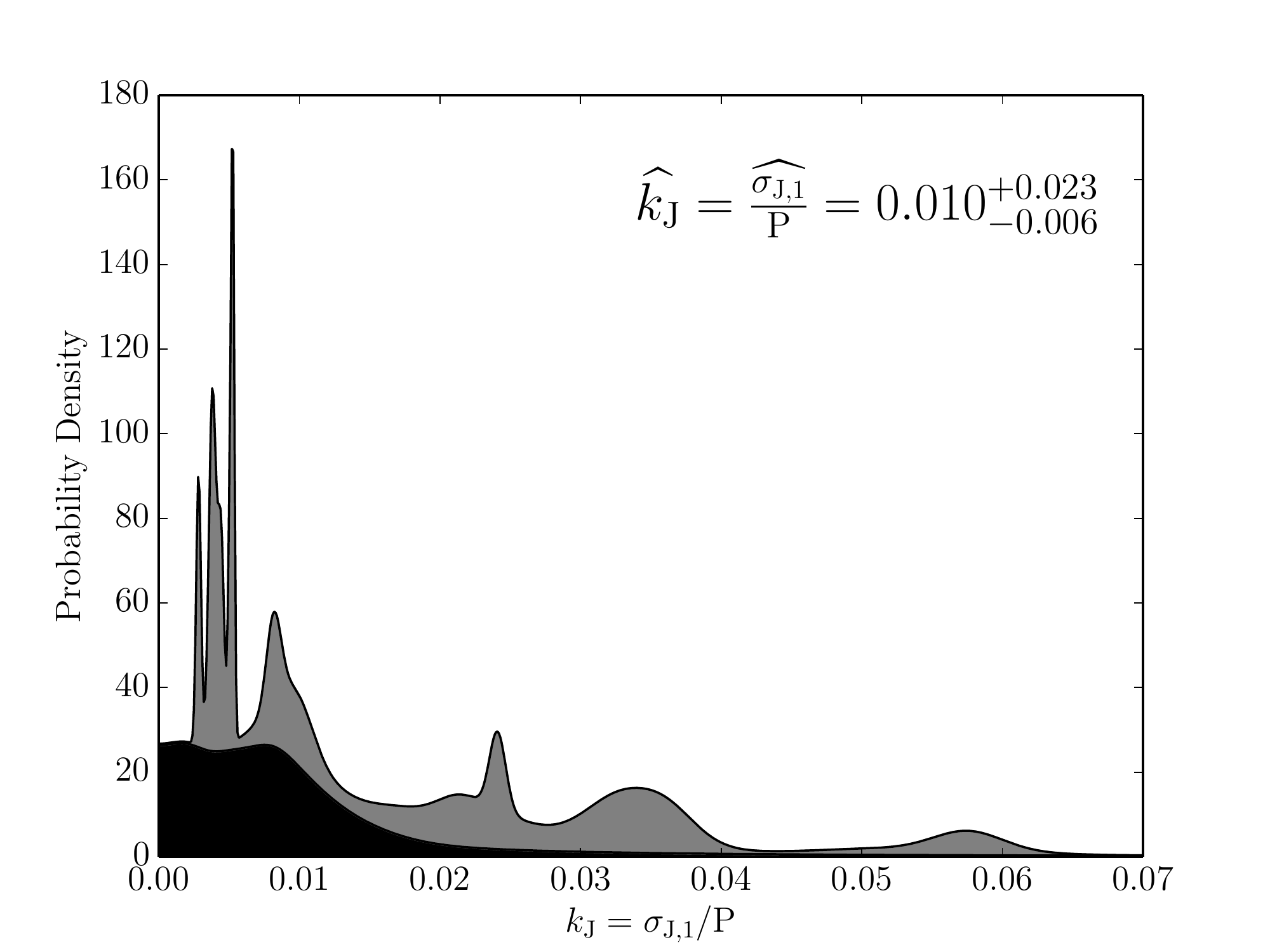}
  \caption{\footnotesize Continuous histogram of the jitter parameter $k_\J = \sigma_{\J,1}/P$. The shaded regions denote the probability density associated with measured values (gray) and upper limits (black) of $k_\J$.}
\label{fig:jitter_distribution}
\end{center}
\end{figure}

Since $\sigma_\DISS\!\ll\!\sigma_\J$ for most pulsars in our sample at 1400~MHz, we can use the likelihood functions $\Like(\sigma_\C)$ computed in the ML analysis (see Eq.~\ref{eq:separable_likelihood}) as a proxy for the likelihood functions $\Like(\sigma_\J)$. In the case of PSR B1937+21, we explicitly set $\Like(\sigma_\J) = \Like(\sqrt{\sigma_\C^2 - \sigma_\DISS^2})$. We ignore PSR J1903+0327 as the upper-limit $\Like(\sigma_\C)$ translates non-trivially to $\Like(\sigma_\J)$. We create a continuous histogram that is the sum of the individual likelihoods $\Like(\sigma_\J)$, shown in Figure~\ref{fig:jitter_distribution}. The black region shows the contributions from upper limit pulsar jitter values and the gray region shows the contributions from measured pulsar values. The median jitter parameter is $\kJ = \sigma_\J / P = 0.010_{-0.006}^{+0.023}$.

\section{Noise Model and Implications for PTA Optimization}
\label{sec:implications}

The noise covariance matrix for short timescales implied by our analysis is
\ba
\mathbf{C}_{\nu\nu',t t'} & = & \delta_{t t'}\left[\delta_{\nu\nu'}\sigma_{\SN}^2(S) + \sigma_\J^2(T)\right] \nonumber \\
& & + \rho_{\DISS,\nu\nu',t t'} \sigma_{\DISS}^2(T),
\label{eq:noise_model}
\ea
where $\delta$ is the Kronecker delta and $\rho_{\DISS,\nu\nu',t t'}$ encapsulates the correlation scales for DISS and we assume that bandwidth is fixed for each receiver. \citet{sod+2014} find that jitter decorrelates over a range of frequencies larger than the total bandwidth of any receiver used in NG9; a decorrelation term is therefore not included in our model. We re-emphasize that $\sigma_{\SN}$ can be calculated directly from the template shape and $\sigma_\J$ is fixed for a given pulsar-frequency combination.

Pulsars dominated by template-fitting errors will see the greatest increase in timing precision from increased integration time and larger bandwidth instrumentation. Wideband timing systems allow for observations of an increased number of scintles and a reduction of $\sigma_{\DISS}$. Therefore, higher DM pulsars, dominated by scintillation noise, will improve in timing quality and will would then become attractive candidates for inclusion into PTAs. By contrast, pulsars dominated by jitter on many epochs do not benefit substantially from wideband timing, though their timing precision will always improve with the increased numbers of pulses observed.

Scintillation monitoring is required in order to characterize the time-varying scintillation parameters, which will not only change $\sigma_\DISS$ but change the PBF over timescales much greater than that of a single epoch. Changes in the PBF will alter pulse shapes and therefore introduce a timing delay into any TOA estimate and contribute to the total white-noise variance.

NG9 uses an empirical, parameterized noise model fit in the timing analysis \citep{NG5BWM,NG9yr,NG9GWB}. For TOAs with an associated error $\sigma_{\SN}$ from template-fitting, the white-noise model is 
\be
\mathbf{C}_{\nu\nu',t t'} = \delta_{t t'}\left[\delta_{\nu\nu'}\left(\mathcal{F}^2 \sigma_{\SN}^2(S) + \mathcal{Q}^2\right) + \mathcal{J}^2\right],
\label{eq:equad}
\ee
where $\mathcal{F}$ (commonly referred to as EFAC) is a dimensionless, constant multiplier to the template-fitting error, $\mathcal{Q}$ (EQUAD) accounts for sources of Gaussian white noise added in quadrature to the template-fitting error, $\mathcal{J}$ (ECORR) accounts for sources of white noise correlated in frequency such as jitter. In NG9, $\mathcal{F} \approx 1$ for all pulsars, to within a factor of 2 for most pulsars. NG9 also fits a red noise model that is negligible on the timescales of a single epoch.

\begin{figure}[t!]
\hspace{-10ex}
\includegraphics[width=0.6\textwidth]{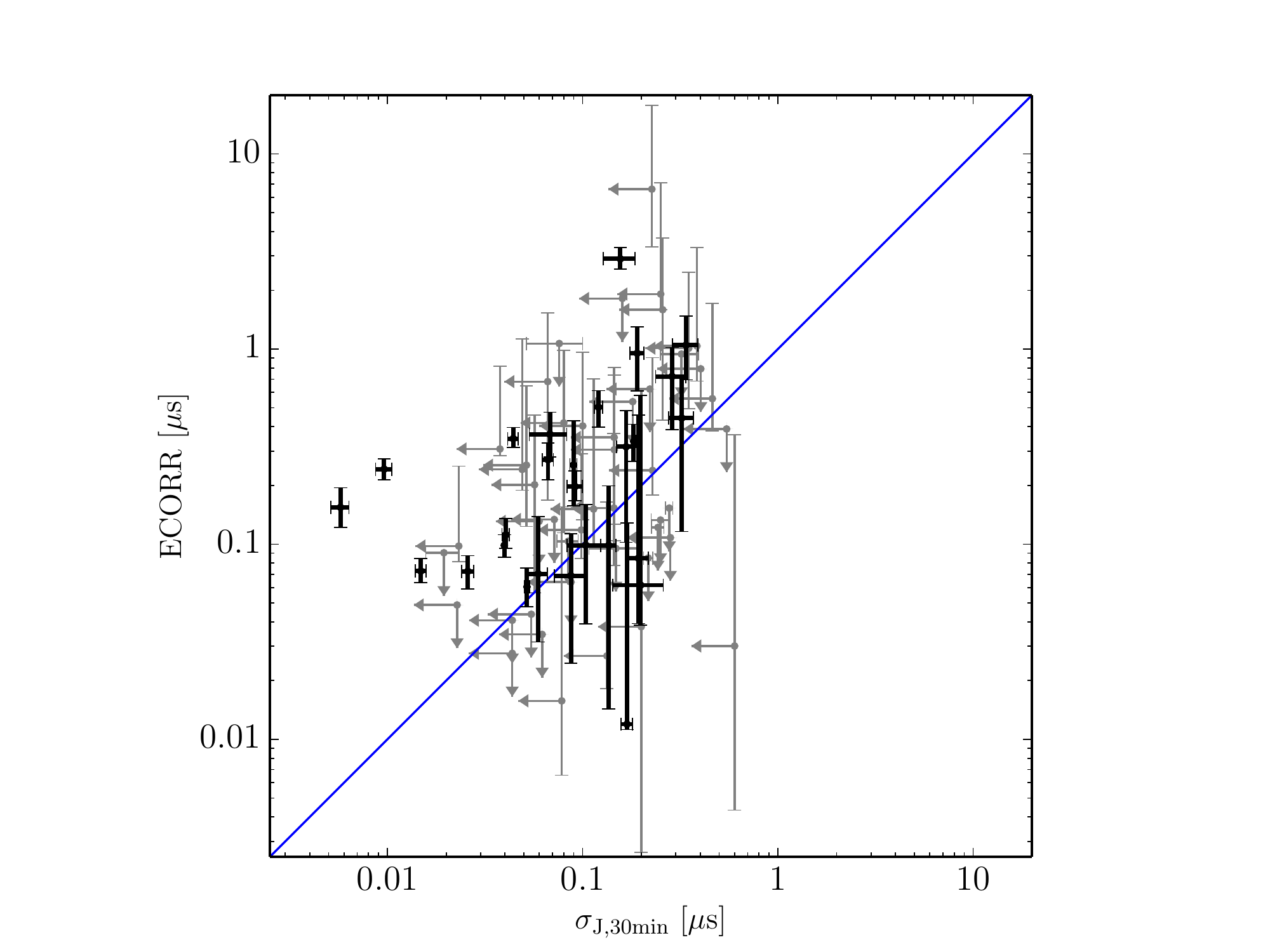}
  \caption{\footnotesize Comparison of ECORR from NG9 with jitter values from this work. Black points denote measurements in both while the gray points mark upper limits in at least one of the values for a given pulsar/receiver combination. The diagonal blue line shows where ECORR equals $\sigma_{\J,30{\rm min}}$.}
\label{fig:ecorr_comparison}
\end{figure}

Figure \ref{fig:ecorr_comparison} shows the comparison between measurements of $\mathcal{J}$ versus $\sigma_{\J,30{\rm min}}$ in black, with the gray points showing values where at least one of the two estimates is an upper limit. We find that $\sigma_{\J,30{\rm min}} \lesssim \mathcal{J}$, which suggests that ECORR is systematically measuring increases in the variance of the residuals, correlated in frequency, beyond pulse jitter. For example, broadband RFI can cause correlations in TOAs measured at different frequencies if unremoved. Replacement of the NG9 empirical white-noise model with our measurements will reduce the number of free parameters in the timing analysis and should improve overall sensitivity to GWs.

\vspace{5ex}

\section{Conclusions}
\label{sec:conclusions}

The short-term white-noise model for pulsar timing is well defined. We have estimated or placed limits on the contributions of the noise model's three white-noise components in the timing residuals of the NANOGrav PTA. The template-fitting errors are consistent with Eq.~\ref{eq:radiometer_noise} and dominate TOA precision for many of the pulsars for many epochs, but scintillation makes jitter important for the higher $\SN$ epochs and TOAs. We find that the template-fitting and jitter errors can be estimated with only pulse $\SN$ as a parameter. The total short-term variance needs contemporaneous measurements of scintillation parameters during observations to properly estimate the time-varying $\sigma_{\DISS}$ contribution. Errors in pulse polarization calibration, or those errors introduced from unremoved RFI, will produce extra variance on short timescales. Long-term observations spanning multiple epochs will have extra variance compared to the short-term model due to a variety of effects that are not included in our analysis.

A large subset of our observed pulsars are jitter-dominated on many epochs and we have measured jitter values for 22 of 37 pulsars. Major improvements in TOA estimation can therefore only be made through increased integration time. For several pulsars, however, $\sigma_\DISS$ is an important if not dominant contribution to the residuals. Wideband timing systems can yield improvements in pulsars with higher DMs such as PSRs B1937+21, J1600$-$3053, J1903+0327, and even moderate-DM pulsars like J2317+1439. Such systems can also improve the average S/N over all epochs, and therefore gains in timing precision can still be made for nearly all of the NANOGrav pulsars.

Jitter appears to be correlated in frequency over each band but not in time. We find that the RMS phase variations from jitter are of order 1\% of the pulse period, though with an extended tail towards higher values of the jitter parameter $k_\J$. Current noise models, such as the one used in NG9, utilize an empirical parameterization that overestimates the RMS jitter. Replacement of model fit parameters with those that can be fixed will ultimately increase sensitivity of the PTA to GWs.

Future telescopes with increased collecting area and sensitivity will become jitter- and DISS-noise dominated. Arrays can therefore be partitioned and pointed at multiple pulsars simultaneously rather than one after another, providing longer integration times for each pulsar and increasing the number of pulses being averaged to reduce the jitter error contribution. The sub-arrays can be partitioned to minimize TOA uncertainty per target pulsar using the formalism outlined here. Wideband timing systems that allow for fine frequency- and time-resolution are needed to fully characterize scintillation structures on a per-epoch basis.

\acknowledgements

{\it Author contributions.} 
MTL developed the mathematical framework, created the modified data set and residuals, undertook the analysis, and prepared the majority of the text, figures, and tables. JMC and SC helped with the development of the framework, the format of figures and tables, and generated some additional text. JAE, DRM, and XS contributed useful discussions regarding the development of the framework. PBD assisted in the initial creation of the modified data set. ZA, KC, PBD, TD, RDF, EF, MEG, GJ, MJ, MTL, LL, MAM, DJN, TTP, SMR, IHS, KS, JKS, and WWZ all ran observations and developed timing models for the NG9 data set. Additional specific contributions are described in \citet{NG9yr}. JAE developed the noise model in NG9 and assisted in the comparison analysis to this work. JMC, SC, GJ, and DJN helped with review of the manuscript.

{\it Acknowledgments.} We would like to thank Michael Jones for useful discussions on statistical methods. The NANOGrav Project receives support from NSF PIRE program award number 0968296 and NSF Physics Frontier Center award number 1430284. NANOGrav research at UBC is supported by an NSERC Discovery Grant and Discovery Accelerator Supplement and the Canadian Institute for Advanced Research. MTL acknowledges partial support by NASA New York Space Grant award number NNX15AK07H. JAE acknowledges support by NASA through Einstein Fellowship grant PF3-140116. Portions of this research were carried out at the Jet Propulsion Laboratory, California Institute of Technology, under a contract with the National Aeronautics and Space Administration. TTP was a student at the National Radio Astronomy Observatory (NRAO) while this project was undertaken. Data for the project were collected using the facilities of the NRAO and the Arecibo Observatory. The NRAO is a facility of the NSF operated under cooperative agreement by Associated Universities, Inc. The Arecibo Observatory is operated by SRI International under a cooperative agreement with the NSF (AST-1100968), and in alliance with the Ana G. M\'{e}ndez-Universidad Metropolitana, and the Universities Space Research Association.

\appendix

\section{Deviations from the Initial Timing Model}
\label{appendix:deviations}

Errors in the initial timing model parameters used for pulse folding and de-dispersion cause effects that can be separated into two related categories: an increase in TOA uncertainties from pulse shape changes on the subintegration timescale $t_{\rm sub}$, and correlated TOA errors over the observation duration $T$. The quadratic fit of initial timing residuals in Eq.~\ref{eq:quadratic_fit} will remove the latter, whereas the former cannot be mitigated after data collection. In part \ref{subsec:smearing} we discuss the non-removable pulse shape changes, in part \ref{subsec:deviations} we discuss the systematic deviations from the initial timing residuals that we remove with our quadratic fit, and in part \ref{subsec:other} we discuss other miscellaneous effects that can cause departures from the initial timing model.

\subsection{Irreversible Pulse Profile Smearing}
\label{subsec:smearing}

{\em 1. Spin period errors:} If the initial folding period is incorrect by an amount $\delta P$, pulse profiles will be smeared by an amount 
\be
\sigma_P \sim \frac{\delta P}{P} t_{\rm sub}. 
\ee
For isolated pulsars, the dominant folding error is due to an error in spin period,
\be
\sigma_P \sim 10~\mathrm{ps}~\left(\frac{\delta P}{10^{-16}~\mathrm{s}}\right) t_{\mathrm{100s}} P_{\rm ms}^{-1},
\ee
where the typical error in the initial folding period is $10^{-16}$~s for pulsars in the NANOGrav data set. Note that periods fit over many years of data are known to much higher precision.

{\em 2. Binary orbit parameter errors:} For binary pulsars, the observed pulse period for low-eccentricity MSPs is Doppler-shifted by an amount \citep{handbook} 
\ba
\sigma_{P_b} & \sim & P \frac{\delta v_\parallel}{c} \sim \frac{2 \pi P}{c} \delta\left(\frac{a \sin i}{P_b} \right) \sim \frac{2\pi P a \sin i}{c P_b} \sqrt{\left(\frac{\delta a}{a}\right)^2 + \left(\frac{\delta (\sin i)}{\sin i}\right)^2 + \left(\frac{\delta P_b}{P_b}\right)^2}\\
& \sim & 72.7~\mathrm{ns}~P_{\mathrm{ms}}~a_{\mathrm{lsec}}~\sin i~P_{b,\mathrm{day}}^{-1} \sqrt{\left(\frac{\delta a}{a}\right)^2 + \left(\frac{\delta (\sin i)}{\sin i}\right)^2 + \left(\frac{\delta P_b}{P_b}\right)^2}
\label{eq:binary_orbit_errors}
\ea
where $a$ is the semi-major axis, $i$ is the inclination angle, and $P_b$ is the binary orbital period, and we assume that the errors in the binary parameters are uncorrelated. The error on these three parameters is much larger than the spin period error, with $\delta(\sin i)/\sin i \sim 10^{-3}$ dominating the other two binary error terms in the NANOGrav initial timing models even when $\sin i$ is well-measured. Therefore, for typical pulsar parameters and when $\sin i$ is measurable, the profile smearing error will be comparable to the spin period error but still negligible. Otherwise, the timing error will be of the order of 10s of nanoseconds.

{\em 3. DM variations:} Differences in the initial timing model DM from the actual DM will cause smearing of pulse profiles. The timing perturbation is roughly the  error in the dispersive delay across a frequency channel \citep{Cordes2002}:
\be
\delta t_{\DM} \simeq 8.3~\mathrm{\mu s}~\delta \DM\;B_{\mathrm{MHz}} \nu_{\mathrm{GHz}}^{-3},
\label{eq:DMerror}
\ee
with $\delta \DM$ measured in units of pc~\cmthree. The typical range in total DM variation in the NANOGrav data set is $\sim 10^{-3}$~pc~\cmthree, which given a 50~MHz channel bandwidth and an observing frequency of 1~GHz, yields a timing perturbation of $\sim 400$~ns. A constant DM over the observation is removed by the term $K(\nu)$ in the timing model fit. Intra-observation DM variations are discussed later in \ref{subsec:deviations}.2.

{\em 4. Polarization calibration errors:} Incorrect gain calibration and summation of the polarization profiles into the intensity profiles will cause pulse shape changes that lead to TOA uncertainties when fitting with a template. The TOA error from gain variation for circularly polarized channels is \citep{ckl+2004}
\be
\delta t_{\rm pol} \sim 1~\mathrm{\mu s} \; \varepsilon_{0.1} \pi_{\rm V,0.1} W_{0.1\mathrm{ms}},
\ee
where $\varepsilon = \delta g/g$ is the fractional gain error, $\pi_{\rm V}$ is the degree of circular polarization, and $W$ is the pulse width. Timing offsets from gain calibration errors will vary slowly with time and will be removed by the quadratic fit discussed in the next sub-section (see for example \ref{subsec:deviations}.3).


\subsection{Systematic Deviations from the Quadratic Fit of Initial Timing Perturbations}
\label{subsec:deviations}

{\em 1. Binary orbit parameter errors:} For pulsars in short-period binary orbits, we will need to fit out higher order terms in $t$ when the period is of order the integration time over the epoch and the binary parameter errors are large. The shortest period binary in NG9 is PSR J0023+0923 with a period of 200~min, nearly seven times longer than the typical total integration time per epoch. The quadratic fit in Eq.~\ref{eq:quadratic_fit} will approximate the sinusoidal variations in TOA offsets introduced by the orbit mis-estimation. The next dominant polynomial term is the cubic term, with error $\sim \sigma_{P_b} (T/P_b)^3$, where $(T/P_b)^3 \sim 0.15^3 \sim 3.4 \times 10^{-3}$ for PSR J0023+0923 and smaller for all other pulsars in the NANOGrav data set. Therefore, using Eq.~\ref{eq:binary_orbit_errors}, the error is negligible.

{\em 2. Ionospheric DM variations:} Changes in DM over short timescales, such as from ionospheric variations, will cause $K(\nu)$ to have time-dependence. The ionospheric DM will vary over the time span of a day due to the changing incident solar flux on a position on the Earth's surface by an amount $\lesssim 3 \times 10^{-5}$~pc~\cmthree~\citep{Lam+2015}. The timing error is approximately the error in the dispersive delay across a frequency channel, given by Eq.~\ref{eq:DMerror}. For a maximum change in DM of $3\times 10^{-5}$~pc~\cmthree~over a 12~hr period, a 1~hr observing length, a 50~MHz channel bandwidth, and an observing frequency of 1~GHz, the timing perturbation is $\approx 1$~ns. Therefore, over the observing span, the assumption that $K(\nu,t) \approx K(\nu)$ holds.


{\em 3. Cross-coupling errors:} Instrumental self-polarization will cause a slow, secular variation in the initial timing residuals when unremoved \citep{ckl+2004}. Cross coupling in the feed will induce a measured false circular polarization $\pi_{\rm V} \simeq 2 \eta^{1/2} \pi_{\rm L}$, where $\eta$ is the voltage cross coupling coefficient and $\pi_{\rm L}$ is a pulsar's degree of linear polarization. While the associated timing errors can be large, errors introduced by the cross-coupling term will cause a slow, secular variation in the residuals as the feed rotates during an observation and will therefore be removed by our quadratic fit. Estimates of these parameters and the induced timing uncertainties will be focused on in future papers.

{\em 4. Rotation Measure (RM) variations:} Faraday rotation from magnetic fields along the pulse propagation path causes both a birefringent TOA delay and the pulse polarization position angle (PPA) to rotate. Changes in the rotation measure ($\RM = \int dl\;n_e B_\parallel$, in units of rad~m$^{-2}$) over short timescales can come from ionospheric variations as with DM. The birefringent delay is given as \citep{Cordes2002}
\be
\delta t_{\RM} = 0.18~{\rm ns}~\RM~\nu^{-3}_{\rm GHz}
\ee
and the change in PPA is \citep{handbook}
\be
\Delta \Psi_{\rm PPA} = \RM~\lambda^2 = 0.09~\RM~\nu^{-2}_{\rm GHz}.
\ee
The RM through the ionosphere is $\sim 1~\mathrm{rad~m^{-2}}$ with $\sim 10\%$ variations on the timescale of 1~hr \citep{S-B+2013} and the birefringent delay is therefore negligible over short timescales. The change in the PPA will cause errors in the polarization calibration that are slowly varying with time and therefore removed by the quadratic fit.

{\em 5. Intrinsic pulsar spin noise:} Rotational instabilities in the pulsar cause deviations from the initial timing model with a steep, power-law noise spectrum over the timescale of years \citep{Cordes2013}. \citet{sc2010} measured spin noise in radio pulsars to scale as $\sigma_{\rm spin} \propto T^{2.0\pm 0.2}$. The pulsar with the largest measured RMS spin noise in the NANOGrav data set is PSR B1937+21, with $\sigma_{\rm spin} \approx 1.5~\mu$s over 10~yr \citep{sc2010,NG9yr}. The RMS on the timescale of 1~hr is $\sim 0.2$~fs and is therefore negligible.

{\em 6. Stochastic GW background:} Like intrinsic pulsar spin noise, GW perturbations will also induce long-term correlations in residuals. However, the RMS timing perturbation from a stochastic GW background of supermassive black hole binaries over 10~yr is on the order of 100~ns \citep{Siemens2013}. The RMS is expected to scale as $\sigma_{\rm GW} \propto T^{5/3}$, and therefore on the timescale of 1~hr the RMS is $\sim 0.6$~fs and is also negligible.

\subsection{Increases in Variance from Other Effects}
\label{subsec:other}

{\em 1. Frequency-dependent DM:} \citet{css2015} describe differences in DM measured at different frequencies due to multipath scattering in the ISM and different volumes of electrons probed. The different $\DM$s as a function of frequency cause differences in the frequency-dependent delays per channel, $K(\nu)$. However, the timescale of refractive variations are weeks or longer, and are therefore this effect is negligible on short timescales.

{\em 2. Mean PBF variations:} As with frequency-dependent DM, the changes in PBFs will occur on a pulsar's refractive timescale and will therefore be negligible on timescale of an hour.


{\em 3. Pulsar mode changes:} Any potential mode changes may cause timing parameter differences from the initial timing model. Pulse profile shapes in our MSPs have not been shown to deviate from the template over the timespan of single observations and any possible epoch-to-epoch mode changes are small and will be removed by our quadratic fit to obtain the short-term timing model (Eq.~\ref{eq:quadratic_fit}).

{\em 4. Transient events:} Giant pulses have been seen in pulsars such as PSR B1937+21 and cause pulse shapes to deviate from the average template shape \citep{Cognard+1996,jap2001,Zhuravlev+2013}. For PSR B1937+21, giant pulses will appear at a rate of approximately 0.5 per 10~s pulse average, which spans $\approx 6400$ pulse periods. Therefore, the giant pulse S/N must be a factor of $\sim \sqrt{6400} \sim 80$ larger than the average single-pulse S/N in order to dominate the template matching fit and significantly alter the estimated TOA. The flux density of the strongest giant pulse in \citet{Zhuravlev+2013} is a factor of $\sim 3$ smaller than the threshold needed to affect the TOA estimation. 

{\em 5. Remaining RFI:} Any remaining RFI in the pulse profiles will introduce unmodeled variance into our analysis. Broadband RFI can cause correlations between residuals that can increase estimates of jitter.

\section{PDF of TOA Errors due to Combined Additive Noise and ISS Modulation}
\label{appendix:pdfs}

The template-fitting error (Eq.~\ref{eq:radiometer_noise}) can be written in the form
\be
\sigma_{\SN} = \sigma_0 \frac{S_0}{S}.
\ee
Again, $S$ is the S/N, proportional to $(S_{\mathrm{PSR}}/\mathrm{SEFD}) \sqrt{2 B T}$, where $S_{\mathrm{PSR}}$ is the pulsar flux density, SEFD is the system equivalent flux density, $B$ is the receiver bandwidth, and $T$ is the total integration time. The subscript `0' is used to denote intrinsic values. We assume that $S_0$ is constant, meaning that both the pulsar flux density and system parameters are also constant (see \S\ref{sec:distributions}).

We describe changes in $S_0$ with a multiplicative gain factor $g$ such that $S = g S_0$. The PDF of the scintillation gains due to DISS with $\niss$ scintles contributing to the measured profile is given by a chi-squared distribution with $2\niss$ degrees of freedom \citep[][Appendix B]{cc1997}:
\be
f_g(g \vert \niss) = \frac{(g\niss)^{\niss}}{g\Gamma(\niss)} e^{-g\niss}\Theta(g).
\label{eq:appendix_scintillation_pdf}
\ee
Unlike DISS, gains from RISS will vary slowly with both time and frequency. For media that follow a Kolmogorov-type electron density wavenumber spectrum with small refractive modulations, DISS and RISS are decoupled in the strong scattering regime. RISS will have a symmetric PDF if focusing is not important and can be approximated with a Gaussian distribution, $f_{g_{\RISS}}(g) = \mathcal{N}(0,\sigma_{\RISS}^2)$  with some correlation time much greater than the observing duration $T$ \citep{Stinebring+2000}. The total gain can be written $g = g_{\DISS} g_{\RISS}$.

We can solve for the PDF of scintillated pulse S/Ns under a change of variables. Eq~\ref{eq:appendix_scintillation_pdf} becomes
\ba
f_S(S \vert \niss) & = & f_g(g \vert \niss) \frac{dg}{dS} \\
& = & f_g\left(\frac{S}{S_0} \middle| \niss\right) \frac{1}{S_0}\\
& = & \frac{(S\niss/S_0)^{\niss}}{S\Gamma(\niss)} e^{-S\niss/S_0}\Theta(S).
\label{eq:g_to_S}
\ea

We can also quantify the distribution of TOA errors, $\Dt$, from scintillation. Errors solely from template fitting in the unscintillated case, $\Dto$,  will be normally-distributed, written as
\be
f_{\Dto}(\Dto) = \mathcal{N}(0,\sigma_{S_0}^2).
\label{eq:appendix_tf_pdf}
\ee
As in Eq~\ref{eq:radiometer_noise}, we rewrite the RMS error is
\ba
\sigma_{\SN} & = & \frac{\Weff}{S \sqrt{\Nphi}}\\
& = & \frac{\Weff}{gS_0 \sqrt{\Nphi}}\\
& = & \frac{\sigma_{S_0}}{g}
\ea
Again, under a change of variables, we can write
\ba
f_{\Dt}(\Dt\vert g) & = & f_{\Dto}(\Dto) \frac{d\Dto}{d\Dt}\\
& = & g f_{\Dto}(g\Dt)
\ea
For brevity, we will write $Z = |\Dt|/\sigma_{\Dto}$. The marginal PDF is then
\ba
f_{\Dt}(\Dt) & = & \int_{-\infty}^\infty dg\;f_g(g) f_{\Dt}(\Dt \vert g)\\
& = & \int_{-\infty}^\infty dg\;g\;f_g(g) f_{\Dto}(g \Dto)\\
& = & \int_{-\infty}^\infty dg\;g\;e^{-g}\;\Theta(g)\;\frac{1}{\sigma_{S_0} \sqrt{2\pi}} \;\exp\left(-\half Z^2 g^2\right)\\
& = & \frac{1}{\sigma_{S_0} \sqrt{2\pi}} \int_0^\infty dg\;g\;e^{-g}\;\exp\left(-\half Z^2 g^2\right)
\ea
From \citet{GR}, Eq.~3.462.1, we have 
\be
\int_0^\infty x^{\alpha-1}e^{-\beta x^2-\gamma x} dx = (2\beta)^{-\alpha/2}\Gamma(\alpha)e^{\gamma^2/(8\beta)}D_{-\alpha}\left(\frac{\gamma}{\sqrt{2\beta}}\right), \mathrm{Re}\;\beta > 0, \mathrm{Re}\;\alpha > 0,
\ee
where $D_n(x) = 2^{-n/2}e^{-x^2/4}H_n(x/\sqrt{2})$ is the Parabolic Cylinder Function defined in terms of the Hermite Polynomial of order $n$, $H_n(x)$. For this calculation, we have $\alpha=\niss+1, \beta=\half Z^2, \gamma=\niss$. Thus, we can write

\ba
f_{\Dt}(\Dt\vert\niss) & = & \frac{1}{\sigma_{S_0} \sqrt{2\pi}} \frac{\niss^{\niss}}{\Gamma(\niss)} Z^{-(\niss+1)} \Gamma(\niss+1)\exp\left({\frac{\niss^2}{4Z^2}}\right) D_{-(\niss+1)}\left(\frac{\niss}{Z}\right)\\
& = & \frac{1}{\sigma_{S_0} \sqrt{2\pi}} \left(\frac{\niss}{Z}\right)^{\niss+1} \exp\left({\frac{1}{4}\left(\frac{\niss}{Z}\right)^2}\right)  D_{-(\niss+1)}\left(\frac{\niss}{Z}\right)\\
& = & \frac{1}{\sigma_{S_0} \sqrt{2\pi}} \left(\frac{\sqrt{2}\niss}{Z}\right)^{\niss+1} H_{-(\niss+1)}\left(\frac{\niss}{\sqrt{2}Z}\right)
\ea
In the case where $\niss = 1$, this reduces to
\be
f_{\Dt}(\Dt\vert\niss) = \frac{1}{\sigma_{S_0} \sqrt{2\pi}} \left(Z^2 - \dfrac{\sqrt{\pi} \exp\left(\frac{1}{2Z^2}\right) \mathrm{erfc}\left(\frac{1}{\sqrt{2}Z}\right)}{\sqrt{2}Z^3}\right).
\ee








\vspace{5ex}




\begin{thebibliography}{27}
\expandafter\ifx\csname natexlab\endcsname\relax\def\natexlab#1{#1}\fi

\bibitem[Antoniadis(2013)]{Antoniadis} Antoniadis, J.~I.\ 2013, Ph.D.~thesis, Univ. of Bonn

\bibitem[Armstrong(1984)]{1984Natur.307..527A} Armstrong, J.~W.\ 1984, 
\nat, 307, 527 

\bibitem[Arzoumanian et al.(2015a)]{NG5BWM} Arzoumanian, Z., Brazier, A., Burke-Spolaor, S., et al.\ 2015, \apj, 810, 150 

\bibitem[Arzoumanian et al.(2015b)]{NG9yr} Arzoumanian, Z., Brazier, A., Burke-Spolaor, S., et al.\ 2015, \apj, 813, 65

\bibitem[Arzoumanian et al.(2015c)]{NG9GWB} Arzoumanian, Z., Brazier, A., Burke-Spolaor, S., et al.\ 2015, arXiv:1508.03024 

\bibitem[Backer et al.(1975)]{brc1975} Backer, D.~C., Rankin, J.~M., \& Campbell, D.~B.\ 1975, \apj, 197, 481 

\bibitem[Blandford et al.(1984)]{1984JApA....5..369B} Blandford, R., 
Romani, R.~W., 
\& Narayan, R.\ 1984, Journal of Astrophysics and Astronomy, 5, 369

\bibitem[Cognard et al.(1996)]{Cognard+1996} Cognard, I., Shrauner, J.~A., Taylor, J.~H., \& Thorsett, S.~E.\ 1996, \apjl, 457, L81 

\bibitem[Coles et al.(2015)]{Coles+2015} Coles, W.~A., Kerr, M., Shannon, R.~M., et al.\ 2015, \apj, 808, 113 

\bibitem[Cordes(2002)]{Cordes2002} Cordes, J.~M.\ 2002, Single-Dish Radio Astronomy: Techniques and Applications, 278, 227 


\bibitem[Cordes(2013)]{Cordes2013} Cordes, J.~M.\ 2013, Classical and Quantum Gravity, 30, 224002 


\bibitem[Cordes \& Chernoff(1997)]{cc1997} Cordes, J.~M., \& Chernoff, D.~F.\ 1997, \apj, 482, 971 

\bibitem[Cordes \& Downs(1985)]{cd1985} Cordes, J.~M., \& Downs, G.~S.\ 1985, \apjs, 59, 343 

\bibitem[Cordes \& Lazio(2002)]{NE2001} Cordes, J.~M., \& Lazio, T.~J.~W.\ 2002, arXiv:astro-ph/0207156 

\bibitem[Cordes \& Rickett(1998)]{cr1998} Cordes, J.~M., \& Rickett, B.~J.\ 1998, \apj, 507, 846 

\bibitem[Cordes \& Shannon(2010)]{cs2010} Cordes, J.~M., \& Shannon, R.~M.\ 2010, arXiv:1010.3785 

\bibitem[Cordes et al.(1990)]{cwd+1990} Cordes, J.~M., 
Wolszczan, A., Dewey, R.~J., Blaskiewicz, M., 
\& Stinebring, D.~R.\ 1990, \apj, 349, 245 

\bibitem[Cordes et al.(2004)]{ckl+2004} Cordes, J.~M., Kramer, 
M., Lazio, T.~J.~W., et al.\ 2004, New A Rev., 48, 1413 

\bibitem[Cordes et al.(2015)]{css2015} Cordes, J.~M., Shannon, 
R.~M., \& Stinebring, D.~R.\ 2015, arXiv:1503.08491 

\bibitem[Craft(1970)]{Craft1970} Craft, H.~D., Jr.\ 1970, PhD~thesis, Cornell Univ.


\bibitem[Demorest(2007)]{Demorest2007} Demorest, P.~B.\ 2007, 
Ph.D.~thesis, Univ. California


\bibitem[Demorest et al.(2010)]{Demorest+2010} Demorest, P.~B., Pennucci, T., Ransom, S.~M., Roberts, M.~S.~E., \& Hessels, J.~W.~T.\ 2010, \nat, 467, 1081 

\bibitem[Demorest et al.(2013)]{Demorest+2013} Demorest, P.~B., Ferdman, R.~D., Gonzalez, M.~E., et al.\ 2013, \apj, 762, 94 

\bibitem[Dolch et al.(2014)]{Dolch+2014} Dolch, T., Lam, M.~T., 
Cordes, J., et al.\ 2014, \apj, 794, 21 

\bibitem[DuPlain et al.(2008)]{drd+2008} DuPlain, R., Ransom, S., Demorest, P., et al.\ 2008, \procspie, 7019, 70191D 

\bibitem[Ford et al.(2010)]{fdr2010} Ford, J.~M., Demorest, P., \& Ransom, S.\ 2010, \procspie, 7740, 77400A 


\bibitem[Foster \& Cordes(1990)]{fc1990} Foster, R.~S., \& Cordes, J.~M.\ 1990, \apj, 364, 123 

\bibitem[Gradshteyn et al.(2007)]{GR} Gradshteyn, I.~S., 
Ryzhik, I.~M., Jeffrey, A., \& Zwillinger, D.\ 2007, Table of Integrals, Series, and Products, Seventh Edition by I.~S.~Gradshteyn, I.~M.~Ryzhik, Alan Jeffrey, and Daniel Zwillinger.~Elsevier Academic Press, 2007.~ISBN 012-373637-4, 

\bibitem[Hassall et al.(2012)]{Hassall+2012} Hassall, T.~E., Stappers, B.~W., Hessels, J.~W.~T., et al.\ 2012, \aap, 543, A66 


\bibitem[Hobbs et al.(2010)]{2010MNRAS.402.1027H} Hobbs, G., Lyne, A.~G., 
\& Kramer, M.\ 2010, \mnras, 402, 1027 


\bibitem[Hotan et al.(2004)]{Hotan+2004} Hotan, A.~W., van Straten, W., \& Manchester, R.~N.\ 2004, PASA, 21, 302 

\bibitem[Jenet et al.(2001)]{jap2001} Jenet, F.~A., Anderson, S.~B., \& Prince, T.~A.\ 2001, \apj, 546, 394 

\bibitem[Keith et al.(2013)]{Keith+2013} Keith, M.~J., Coles, W., 
Shannon, R.~M., et al.\ 2013, \mnras, 429, 2161

\bibitem[Kramer(1998)]{Kramer1998} Kramer, M.\ 1998, \apj, 509, 856 

\bibitem[Kramer et al.(1998)]{Kramer+1998} Kramer, M., Xilouris, K.~M., Lorimer, D.~R., et al.\ 1998, \apj, 501, 270 


\bibitem[Lam et al.(2015)]{Lam+2015} Lam, M.~T., Cordes, J.~M., 
Chatterjee, S., et al.\ 2015, arXiv:1512.02203 


\bibitem[Lambert \& Rickett(1999)]{lr1999} Lambert, H.~C., \& Rickett, B.~J.\ 1999, \apj, 517, 299 

\bibitem[Lazaridis et al.(2009)]{Lazaridis+2009} Lazaridis, K., Wex, N., Jessner, A., et al.\ 2009, \mnras, 400, 805 

\bibitem[Lentati et al.(2014)]{Lentati+2014} Lentati, L., Hobson, M.~P., \& Alexander, P.\ 2014, \mnras, 444, 3863 

\bibitem[Levin et al.(2016)]{Levin+2016} Levin, L., McLaughlin, M.~A., Jones, G., et al.\ 2016, arXiv:1601.04490 

\bibitem[Liu et al.(2011)]{Liu+2011} Liu, K., Verbiest, J.~P.~W., Kramer, M., et al.\ 2011, \mnras, 417, 2916 

\bibitem[Liu et al.(2012)]{Liu+2012} Liu, K., Keane, E.~F., Lee, K.~J., et al.\ 2012, \mnras, 420, 361 


\bibitem[Lorimer \& Kramer(2012)]{handbook} Lorimer, D.~R., \& Kramer, M.\ 2012, Handbook of Pulsar Astronomy, by D.~R.~Lorimer , M.~Kramer, Cambridge, UK: Cambridge University Press, 2012,  

\bibitem[Lyne et al.(2013)]{Lyne+2013} Lyne, A., Graham-Smith, F., Weltevrede, P., et al.\ 2013, Science, 342, 598 


\bibitem[McLaughlin(2013)]{McLaughlin2013} McLaughlin, M.~A.\ 2013, Classical and Quantum Gravity, 30, 224008 

\bibitem[Pennucci et al.(2014)]{pdr2014} Pennucci, T.~T., Demorest, P.~B., \& Ransom, S.~M.\ 2014, \apj, 790, 93 

\bibitem[Perera et al.(2010)]{Perera+2010} Perera, B.~B.~P., McLaughlin, M.~A., Kramer, M., et al.\ 2010, \apj, 721, 1193

\bibitem[Phillips \& Wolszczan(1992)]{pw1992} Phillips, J.~A., \& Wolszczan, A.\ 1992, \apj, 385, 273 

\bibitem[Pilia et al.(2015)]{Pilia+2015} Pilia, M., Hessels, J.~W.~T., Stappers, B.~W., et al.\ 2015, arXiv:1509.06396 

\bibitem[Rickett(1990)]{r90}    
Rickett, B.~J.\ 1990, \araa, 28, 56

\bibitem[Shannon \& Cordes(2010)]{sc2010} Shannon, R.~M., \& Cordes, J.~M.\ 2010, \apj, 725, 1607 

\bibitem[Shannon \& Cordes(2012)]{sc2012} Shannon, R.~M., \& Cordes, J.~M.\ 2012, \apj, 761, 64 

\bibitem[Shannon et al.(2014)]{sod+2014} Shannon, R.~M., Os{\l}owski, S., Dai, S., et al.\ 2014, \mnras, 443, 1463 

\bibitem[Shao \& Wex(2013)]{sw2013} Shao, L., \& Wex, N.\ 2013, Classical and Quantum Gravity, 30, 1650

\bibitem[Siemens(2013)]{Siemens2013} Siemens, X.\ 2013, Classical and Quantum Gravity, 30, 224015 

\bibitem[Sotomayor-Beltran et 
al.(2013)]{S-B+2013} Sotomayor-Beltran, C., Sobey, C., Hessels, J.~W.~T., et al.\ 2013, \aap, 552, A58 

\bibitem[Stinebring et al.(2000)]{Stinebring+2000} Stinebring, D.~R., Smirnova, T.~V., Hankins, T.~H., et al.\ 2000, \apj, 539, 300 

\bibitem[Taylor(1992)]{Taylor1992} Taylor, J.~H.\ 1992, Royal 
Society of London Philosophical Transactions Series A, 341, 117 

\bibitem[Turin(1960)]{Turin1960} Turin, G.~L.\ 1960, IRE Transactions on Information Theory, IT-6, 3, 311-329

\bibitem[van Haasteren et al.(2009)]{vH+2009} van Haasteren, R., Levin, Y., McDonald, P., \& Lu, T.\ 2009, \mnras, 395, 1005 

\bibitem[van Straten et al.(2012)]{vS+2012} van Straten, W., Demorest, P., \& Oslowski, S.\ 2012, Astronomical Research and Technology, 9, 237 

\bibitem[Verbiest et al.(2009)]{Verbiest+2009} Verbiest, J.~P.~W., Bailes, M., Coles, W.~A., et al.\ 2009, \mnras, 400, 951 

\bibitem[Will(2014)]{Will2014} Will, C.~M.\ 2014, Living Reviews in Relativity, 17,  

\bibitem[Yan et al.(2015)]{YanZ+2015} Yan, Z., Shen, Z.-Q., Wu, X.-J., et al.\ 2015, \apj, 814, 5 

\bibitem[Zhu et al.(2015)]{Zhu+2015} Zhu, W.~W., Stairs, I.~H., Demorest, P.~B., et al.\ 2015, \apj, 809, 41 

\bibitem[Zhuravlev et al.(2013)]{Zhuravlev+2013} Zhuravlev, V.~I., Popov, M.~V., Soglasnov, V.~A., et al.\ 2013, \mnras, 430, 2815 

\end{thebibliography}
\end{document}